\documentclass{cplslarge}%
\usepackage[authoryear]{natbib}

\usepackage{color}
\definecolor{red}{rgb}{1,0,0}
\definecolor{black}{rgb}{0,0,0}
\definecolor{blue}{rgb}{0,0,1}
\definecolor{green}{rgb}{0,0.7,0}

\usepackage[ddmmyyyy]{datetime}
\usepackage{amssymb}

\usepackage{makeidx}
\makeindex

\begin{document}
\renewcommand{\thechapter}{\arabic{chapter}}    
\setcounter{chapter}{6}
\author[Sicardy \&\ at al.]{bruno~sicardy, maryame~el~moutamid, alice~c.~quillen, paul~m.~schenk, mark~r.~showalter, and~kevin~walsh}

%

\chapter{Rings beyond the giant planets}

\section{Introduction}

Until 2013, only the giant planets were known to host ring systems.
In June 2013, a stellar occulation revealed the presence of narrow and dense
rings around Chariklo, a small Centaur object that orbits between Saturn and Uranus.
Meanwhile, the Cassini spacecraft revealed evidence for the possible past presence 
of rings around the Saturnian satellites Rhea and Iapetus.
Mars and Pluto are expected to have tenuous dusty rings, though they have so far evaded detection. %
More remotely, transit events observed around a star in 2007 may have revealed
for the first time exoplanetary rings around a giant planet orbiting that star.

So, evidence is building to show that rings are more common features
in the universe than previously thought.
Several interesting issues arise from the discovery (or suspicion)
of the new ring systems described in this chapter.
One of them is to assess how universal is the physics governing rings,
in spite of large differences in size, age and origin.
In other words, 
do rings obey some common, fundamental processes, 
or are their similarities just apparent and stemming from very different mechanisms?
Another interesting question is what those ring systems tell us about the 
origin, evolution and physical properties of the bodies they encircle. 
As such, rings may be of precious help to better understand 
the formation of satellites and planets, not only in our own solar system, 
but also among extrasolar worlds.
We will return to those considerations in the concluding remarks of this chapter,
after reviewing recent ring system discoveries.

\section{Dense rings around the small Centaur object Chariklo}

\begin{figure}[!h]
\figurebox{80mm}{}{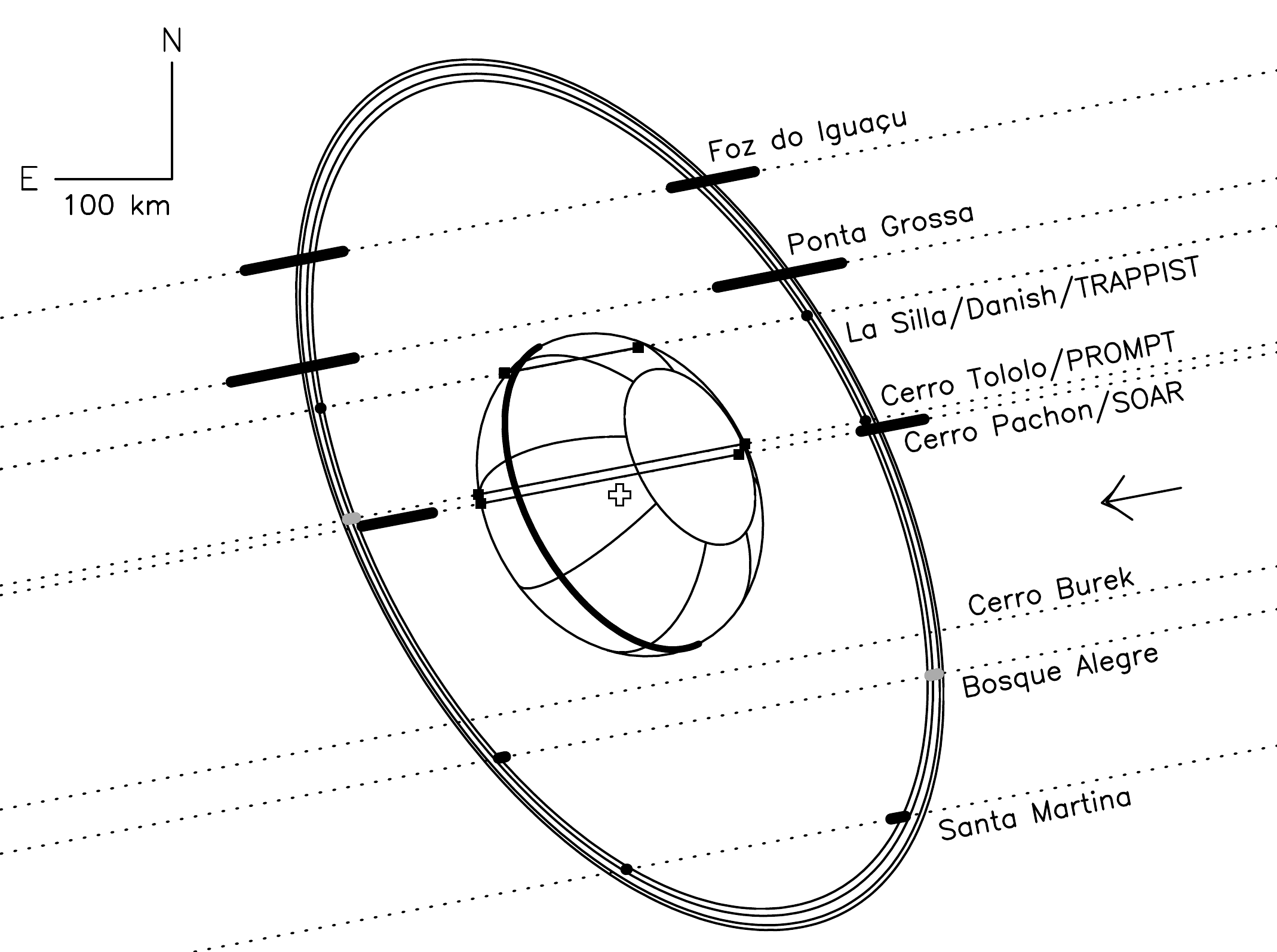}
\caption[The discovery of Chariklo's rings]{%
The discovery of Chariklo's rings during the June 3, 2013 stellar occultation.
The dotted lines show the trajectories of the star 
in the plane of the sky relative to Chariklo, 
as seen from eight stations in Brazil, Argentina and Chile
(the arrow indicates the direction of motion).
The occultation by the main body was observed along the three black segments - or ``chords"~-
near the center of the plot.
Beside these detections, secondary events were observed somewhere inside the black, thick intervals, 
most of them unresolved in time.
Some segments are longer because of the longer integration time used at the corresponding stations,
hence their larger uncertainties in position.
%
The two gray segments along the Bosque Alegre and Cerro Tololo chords correspond to 
dead-times (due to image readouts) during acquisition,  leading to \textit{non} detections of the ring, 
but still providing constraints on the ring location. 
%
The rings were not detected at Cerro Burek due to a large integration time at low signal-to-noise ratio.  
The size, shape and orientation of the inner, denser ring (C1R) is obtained though
an elliptical fit to the black and gray segments, weighted with their respective uncertainties. 
The outer, fainter ring (C2R) was resolved only 
in the best-sampled light curve of the Danish telescope at La Silla (Fig.~\ref{chap7_fig_danish}).
Its orbit has been reconstructed assuming that C1R and C2R are concentric.
Adapted from \cite{chap7_bra14}.
}%
\label{chap7_fig_ring_geom}
\end{figure}

In June 2013, narrow, sharply confined dense rings were discovered around 
the small Centaur object (10199) Chariklo.
This asteroid-like object became the first body of the solar system,
other than the giant planets, known to possess rings.
Meanwhile, those rings resemble some of the sharply defined features observed around Saturn or Uranus 
(Figs.~\ref{chap7_fig_ring_geom}-\ref{chap7_fig_ring_prof}), 
suggesting some common dynamics.
 
Centaurs are small objects (diameters less than about 250~km) 
with perihelion beyond Jupiter's orbit (5.2 AU) and 
semi-major axis inside of Neptune (30.0 AU).
They were originally Trans-Neptunian Objects (TNO's) that have been scattered 
by gravitational tugs from Neptune or Uranus \citep{chap7_gla08}.

\begin{table*}
\caption{Chariklo main body physical parameters}{\tabcolsep5.5pt%
\begin{tabular}{@{}ll@{}}			
\toprule%
Semi-major axis, period$^{a}$         & 15.79 AU,  62.71 yrs \\
Eccentricity, inclination$^{a}$          & 0.1716, 23.37 deg    \\
Perihelion-aphelion$^{a}$               & 13.08--18.50   AU   \\\hline
Equivalent radius$^{b,c}$                & $R_{\rm equiv}= 119 \pm 5$  km \\
Visible geometric albedo$^{b}$      & $p_V= 0.042 \pm 0.005$ \\
Synodical rotation period$^{b}$      & $P_C= 7.004 \pm 0.036$~h     \\\hline
Mass$^{d}$                                         & $\sim 10^{19}$~kg \\\hline
Surface composition$^{e}$                 &  $\sim 60$\% amorphous carbon \\ 
                                                             &  $\sim 30$\% silicates, $\sim 10$\% organics \\ 
\botrule
\end{tabular}}
\begin{tabnote}
$^{a}$\cite{chap7_des15a,chap7_des15b}.
$^{b}$\cite{chap7_for14}.
$^{c}$The radius of a spherical body that presents the same apparent surface area as the actual body.
$^{d}$Order of magnitude estimate, using $R_{\rm equiv}$ above and assuming an icy body.
$^{e}$\cite{chap7_duf14b}.
\end{tabnote}
\label{chap7_tab_ck_phys_param}
\end{table*}

Chariklo was discovered in February 1997 \citep{chap7_sco97}
and is the largest Centaur known to date, with a diameter of about 240~km.
Its very low geometric albedo (about 4\%, see Table~\ref{chap7_tab_ck_phys_param}) 
makes it one of the darkest objects of the solar system.
It moves close to a 4:3 mean-motion resonance with Uranus, its main perturber.
Dynamical studies indicate that Chariklo 
has been captured in its present orbital configuration some 10 Myr ago, and that 
the half-life time of its unstable current orbit is about 10 Myr \citep{chap7_hor04},
a very short timescale compared to the age of the solar system.

Year-scale photometric \citep{chap7_bel10} and spectroscopic \citep{chap7_gui11} 
variations of Chariklo were tentatively attributed to transient periods of cometary activity. 
As discussed below, those variations can be naturally explained by the presence of a flat, 
partially icy ring system observed at various aspect angles, 
so that no cometary activity is necessary to explain this behavior.
Actually, we will see that no dust or gas production has been detected so far around Chariklo.

Meanwhile, Chiron (another Centaur similar in size to Chariklo)
is also surrounded by narrowly confined material whose interpretation is still debated.
%
This shows that material around Centaurs or other small bodies
may be more common than previously thought.

In the following sub-sections, the term ``Chariklo" will apply to the central body only,
while ``Chariklo's system" will denote the entire set Chariklo plus its rings.

\subsection{The discovery of Chariklo's rings}
%

\begin{figure}
\figurebox{80mm}{}{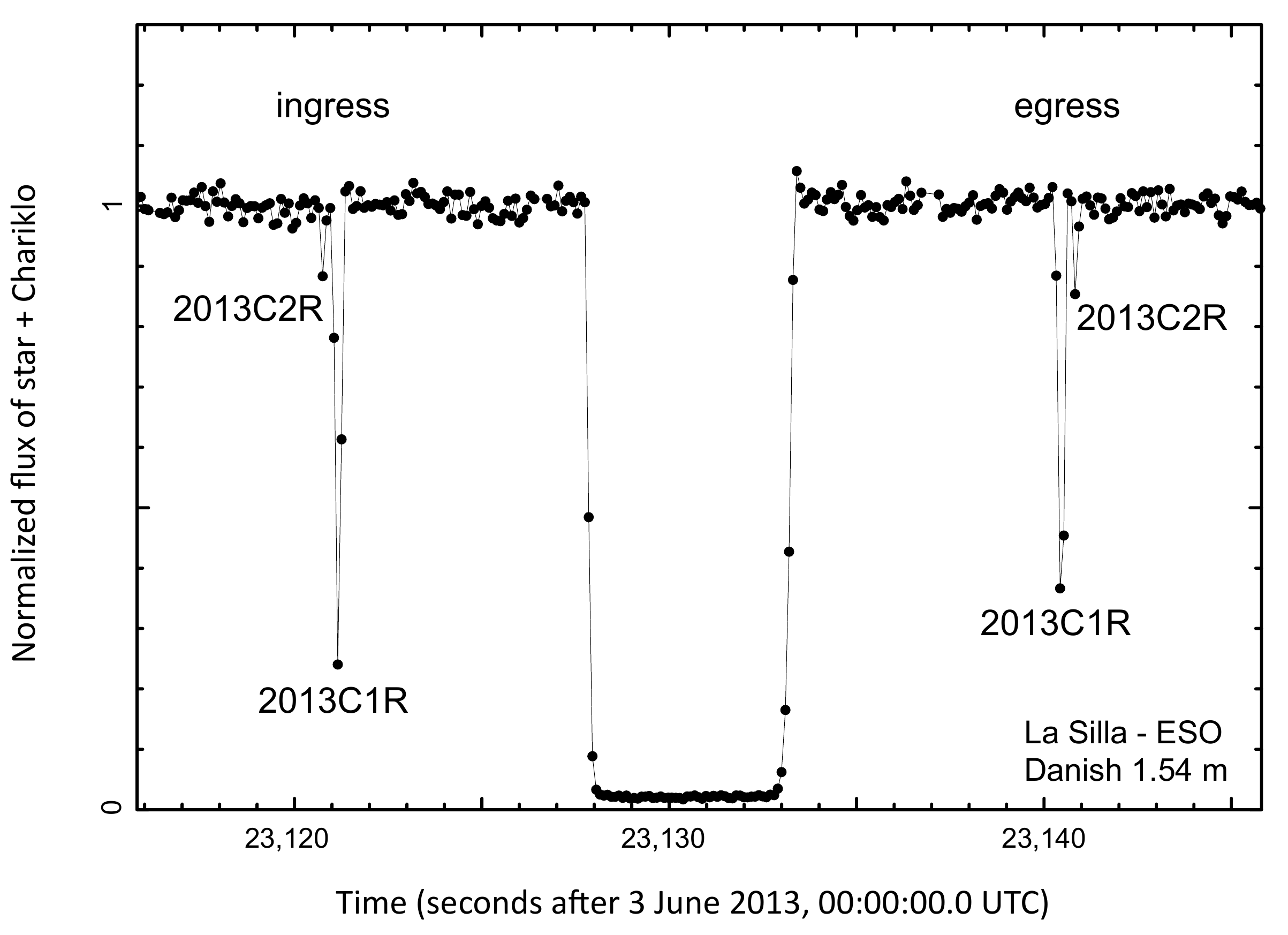}
\caption[Chariklo's ring system detection]{%
Plot of the stellar flux vs. time, as observed from the Danish telescope at La Silla,
during the June 3, 2013 occultation.
This is the best sampled at highest signal-to-noise ratio light curve among the various
sites involved in this observation (Fig.~\ref{chap7_fig_ring_geom}).
It shows a $\sim 5.3$~s central drop, corresponding to the blocking of the stellar
flux by Chariklo's main body.
The two symmetric events on each side are caused by the rings. 
In fact, each event is resolved into a main (C1R) and fainter (C2R) rings 
separated by an essentially empty gap.
Adapted from \cite{chap7_bra14}.
}%
\label{chap7_fig_danish}
\end{figure}

\index{Chariklo's rings, discovery}
Chariklo's rings were discovered during a stellar occultation,
which occurs when an object passes in front of a star, 
blocking its flux for some seconds (Fig.~\ref{chap7_fig_danish}).
Such an event was monitored on June 3, 2013 from various sites in 
Brazil, Uruguay, Argentina and Chile, 
see Fig.~\ref{chap7_fig_ring_geom} and \cite{chap7_bra14}. 
This was the first successful Chariklo occultation ever observed.
This event was one of a number aimed at
characterizing the sizes, shapes and surroundings of  TNO's and Centaurs \citep{chap7_ass12,chap7_cam14},
and monitoring Pluto's atmosphere \citep{chap7_ass10}.
In the case of Chariklo, 
a further incentive was the search for surrounding (possibly cometary) material, 
as both sharp and diffuse secondary events were detected in 1993 and 1994 during stellar occultations 
by its sibling Chiron \citep{chap7_ell95,chap7_bus96}, see below.

Narrow secondary events were indeed detected during the June 2013 
event (Figs.~\ref{chap7_fig_ring_geom} and \ref{chap7_fig_danish}),
but it became rapidly clear that they could not be interpreted 
as collimated, radial  cometary jets ejected from Chariklo's surface
because their geometries were mutually inconsistent with that interpretation.
Instead, the ring interpretation was the simplest, although surprising, explanation 
for all the observed secondary events.
All the detections were in fact consistent with the presence of two rings: 
an inner, denser ring, 2013C1R (C1R for short), orbiting at some 
390~km from Chariklo's center, 
15~km inside another, more tenuous outer ring, C2R 
(Fig.~\ref{chap7_fig_ring_geom}, Table~\ref{chap7_tab_ring_phys_param}).
There were several arguments in favor of the ring interpretation drawn from those observations: 
\begin{itemize}
\item{%
Although most of the stations appearing in Fig.~\ref{chap7_fig_ring_geom} 
did not resolve the rings, their equivalent widths $W_e$
(which measure the amount of material contained in the rings, see Chapter~4)
were essentially the same for all events. 
Such coincidence is hard to reconcile with a set of independent 
cometary jets going in different directions.
}%
\item{%
A flat ring system offers a natural explanation for Chariklo's long-term photometric variations
(see \citealt{chap7_bel10} and Fig.~\ref{chap7_fig_ring_photom}). 
Those variations merely reflect the changing
ring aspect as Chariklo and Earth revolve  around the Sun.
}%
\item{%
The ring interpretation also offers a simple explanation for the appearance and disappearance of the
2.2~$\mu$m water ice band in Chariklo's spectra (\citealt{chap7_gui11} and Fig.~\ref{chap7_fig_spec}). 
Again, the changing ring geometry causes the disappearance and reappearance of the ice band, 
showing by the same token that the rings do contain water ice.
}%
\end{itemize}

Another occultation observed on April 29, 2014 
fully confirmed the ring interpretation drawn from the June 2013 discovery, 
and revealed finer structures in the main ring C1R, 
see Fig.~\ref{chap7_fig_ring_prof} and the associated discussion.
Other occultations revealed either the main body only or the rings alone, 
but with lower signal-to-noise ratios or insufficient resolution 
to reveal ring sub-structures (B\'erard et al. 2016 and Leiva et al. 2016, in preparation).

\begin{table*}
\caption{Chariklo's rings physical parameters}
{\tabcolsep5.5pt%
\begin{tabular}{@{}llll@{}}		
\toprule%
                                                            & Radius$^{a}$            & Radial width                     & Normal optical depth \\\hline
Ring C1R                                          & $390.6\pm3.3$~km  & $4.8<W<7.1$~km$^{b}$ & average $\tau_N \sim 0.4$$^{c}$ \\
Ring C2R                                          & $404.8\pm3.3$~km  &  $W \sim$ 1-3~km            & $\tau_N \sim$ 0.1 \\\hline
\multicolumn{2}{@{}l}{Gap between C1R and C2R$^{a}$}      & $8.7\pm0.4$~km       & $< 0.004$   \\\hline
Pole position$^{a}$ & \multicolumn{3}{l}{$\alpha_p=$~10 h 05 min$\pm$ 02 min, $\delta_p=$ +41$^{\circ}$ 29'$\pm$13' (equatorial J2000)} \\\hline                                                                            
Visible reflectivity$^{d}$        & $(I/F)_V= 0.07 \pm 0.01$  & & \\\hline
Surface composition$^{d}$   & \multicolumn{3}{l}{20\% water ice, 40-70\% silicates, 10-30\% tholins,} \\
                                                  & \multicolumn{3}{l}{small quantities of amorphous carbon} \\                                                                     
\botrule
\end{tabular}
}%
\begin{tabnote}
$^{a}$From \cite{chap7_bra14}, assuming circular rings.
$^{b}$Smallest and largest widths observed during the June 3, 2013 and April 29, 2014 stellar occultations 
\citep[and B\'erard et al. 2016, in preparation]{chap7_sic14}.
$^{c}$ With some some opaque parts, see text and Fig.~\ref{chap7_fig_ring_prof}.
$^{d}$\cite{chap7_duf14b}.
\end{tabnote}
\label{chap7_tab_ring_phys_param}
\end{table*}

\subsection{Physical properties}
\index{Chariklo's rings, physical properties}

\subsubsection{Orbit}
The secondary events shown around Chariklo in Fig.~\ref{chap7_fig_ring_geom} 
constrain the apparent shape, size and orientation of the main ring C1R projected in the sky plane
(the case of the more tenuous, nearby ring C2R is considered in a second step, 
as mentioned in the caption of Fig.~\ref{chap7_fig_ring_geom}).

The simplest model for ring C1R is that of an ellipse with one focus at Chariklo's center of mass.
However, we do not know a priori the ring pole position, nor its apse orientation.
Moreover, we do not have enough occulting chords across the main body 
from the observation of June 2013 (nor from other later ones, to date) to determine Chariklo's 
center position relative to the main ring (Fig.~\ref{chap7_fig_ring_geom}).

Thus, one has to make the simpliflying assumption that Ring C1R is circular, 
with opening angle $B$ and position angle $P$ as seen from Earth.
An elliptical fit to the secondary events of  Fig.~\ref{chap7_fig_ring_geom} then provides 
the center of the ellipse, 
its apparent semi-major and semi-minor axes $a'$ and $b'$ (projected in the sky plane)
and its position angle. 
Note that in the circular assumption, $|\sin(B)|=b'/a'$.

%

%
%

The elliptical fit displayed in Fig.~\ref{chap7_fig_ring_geom} allows two alternative
ring pole positions, depending on which part of the ring is ``in front" the sky plane.
This ambiguity can be solved by considering Chariklo's photometric evolution over time.
The pole position adopted here (Table~\ref{chap7_tab_ring_phys_param}) 
predicts that the rings were observed edge-on in 2008,
in agreement with Chariklo's system photometric behavior (Fig.~\ref{chap7_fig_ring_photom}).
Conversely, the alternative solution predicts an edge-on configuration in 1994 that
is out of phase compared to the observed behavior.
Moreover, the solution adopted here was confirmed during another Chariklo stellar occultation 
observed on April 29, 2014 
(B\'erard et al. 2016, in preparation).
Defining the ring pole direction as being parallel to the ring angular momentum,
there is a further ambiguity since two opposite orbital motions are possible, 
that correspond to opposite values of $B$.
We have arbitrarily chosen $B>0$ in Table~\ref{chap7_tab_ring_phys_param}.
This choice may be revised in the future (in favor of the opposite pole) 
if the particles orbital can be determined.

\begin{figure}
\centerline{
\figurebox{70mm}{}{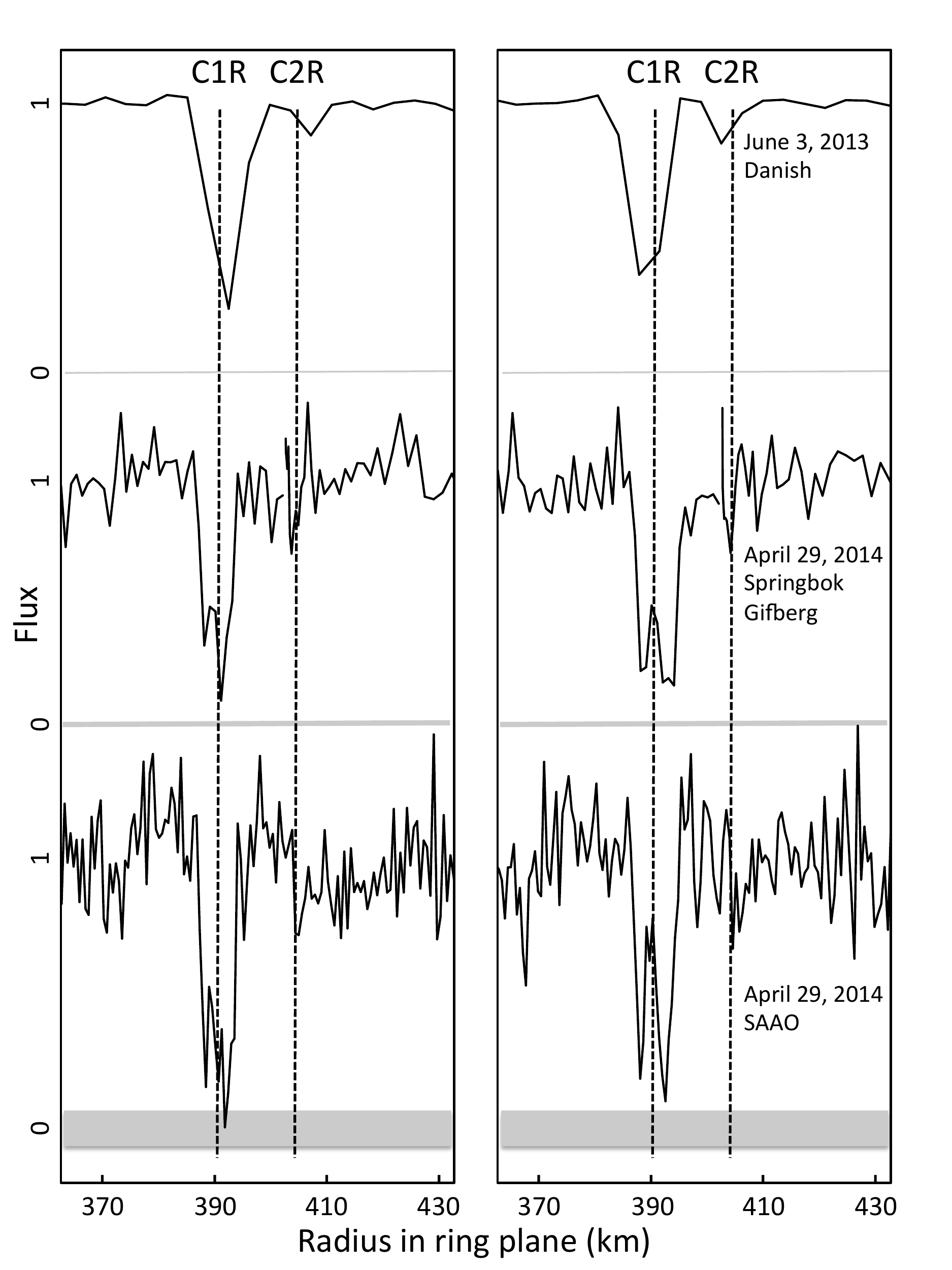}
}
\caption[Chariklo's ring profiles]{%
Chariklo's ring radial profiles derived from the 
June~3, 2013 stellar occultation (Danish telescope, top), the 
April~29, 2014 combined event at Springbok and Gifberg stations (middle), 
and the same event at the South African Astronomical Observatory (SAAO, bottom).
The gray boxes correspond to the zero stellar fluxes (complete star disappearance),
where its thickness represents photometric uncertainties, while 
unity corresponds to the full, unocculted stellar flux. 
The horizontal axis is the distance to Chariklo's center, measured in the plane of the
rings, using the orientation given in Table~\ref{chap7_tab_ring_phys_param}.
This orientation is derived \it assuming \rm that the rings are circular, so that
this plot cannot be used to assess or put upper limit on the ring eccentricities.
The vertical dotted lines are the ring radii adopted in Table~\ref{chap7_tab_ring_phys_param}.
Due to the different acquisition rates and viewing geometries, 
the light curves have radial samplings of 3.6, 1.0 and 0.57~km 
par data point in the top, middle and bottom panels, respectively.
The best resolved profiles are eventually diffraction-limited
at the Fresnel scale limit (about 0.8~km).
They show sharp edges and
a W-shaped structure in the main ring C1R, 
as well as a width variation for ring C1R.
%
}%
\label{chap7_fig_ring_prof}
\end{figure}

The best light curve obtained during the June 3, 2013 occultation shows that there are actually two rings. 
The dense ring C1R is flanked by a more tenuous outer ring, C2R 
(Fig.~\ref{chap7_fig_danish}). 
All the other instruments used during the discovery observation
did not have enough time resolution to separate the two rings,
but they were clearly resolved again during the April 29, 2014 event 
(Fig.~\ref{chap7_fig_ring_prof}). 
The orbital parameters for C1R and C2R, resulting from the fit of Fig.~\ref{chap7_fig_ring_geom},
are listed in Table~\ref{chap7_tab_ring_phys_param}.
It is assumed here that C2R has also a circular orbit, 
14.2~km outside C1R and concentric with it, 
as derived by \cite{chap7_bra14}.

\subsubsection{Fine structure}

From the June 3, 2013 discovery observations,
no material was detected in the gap between C1R and C2R up to a normal optical depth of about 0.004
\citep{chap7_bra14}.
Those observations did not reveal structures inside C1R and C2R, 
due to insufficient time resolution (Fig.~\ref{chap7_fig_danish}).
However, data obtained at higher rate during another event (April 29, 2014) 
revealed a double-dip structure inside ring C1R, 
while no structure has been identified so far in the shallower C2R profiles
(Fig.~\ref{chap7_fig_ring_prof}).

The densest parts of C1R are consistent with opaque material concentrated at very sharp edges.
The main smoothing effect of the April 29, 2014 profiles is Fresnel diffraction, 
which amounts to about  0.8~km when projected at the ring
(the finite stellar diameter being negligible for that event).
At this scale, one cannot resolve the edges, 
as the occultation profiles are compatible with infinitely sharp boundaries 
(B\'erard et al. 2016, in preparation).

So far, only eight C1R profiles obtained in  2013 and  2014 could provide
an estimation the ring radial width $W$ (projected in the ring plane), the other profiles
having insufficient resolution to do so. 
The width $W$ shows significant variations between 4.8 and 7.1~km
(\citealt{chap7_sic14} and Table~\ref{chap7_tab_ring_phys_param}),
with dynamical implications that are discussed later.

For the unresolved profiles, it is possible to estimate the equivalent width
$W_{e(1+2)} = (W_{1} \cdot p_{N1} + W_{2} \cdot p_{N2})$ of the global  ring system C1R+C2R, 
where $W_i$ and $p_{Ni}$ denote the physical width and normal opacitiy 
of each component, respectively.
For a monolayer ring, $W_e$ is a measure of the amount of material 
contained along a radial cut of that ring. 
It can be viewed as the width of an opaque monolayer ring that would
block the same amount of light as the observed ring 
(see \citealt{chap7_ell84} and Chapter~4). 

The fifteen or so C1R+C2R occultation profiles obtained so far provide
a consistent value close to $W_{e(1+2)} \sim 2$~km 
(with typical dispersion of $\sim 1$~km), 
while the six profiles where C2R can be resolved from C1R provide $W_{e2} \sim 0.25$~km 
(with dispersion $\sim 0.05$~km),
with no significant azimuthal variations.
Thus, the system C1R+C2R does not show appreciable azimuthal variations in $W_e$ and 
C2R appears to contain $\sim$ 10 times less material than C1R.

\subsubsection{Photometry}

\begin{figure}
\centerline{
\figurebox{80mm}{}{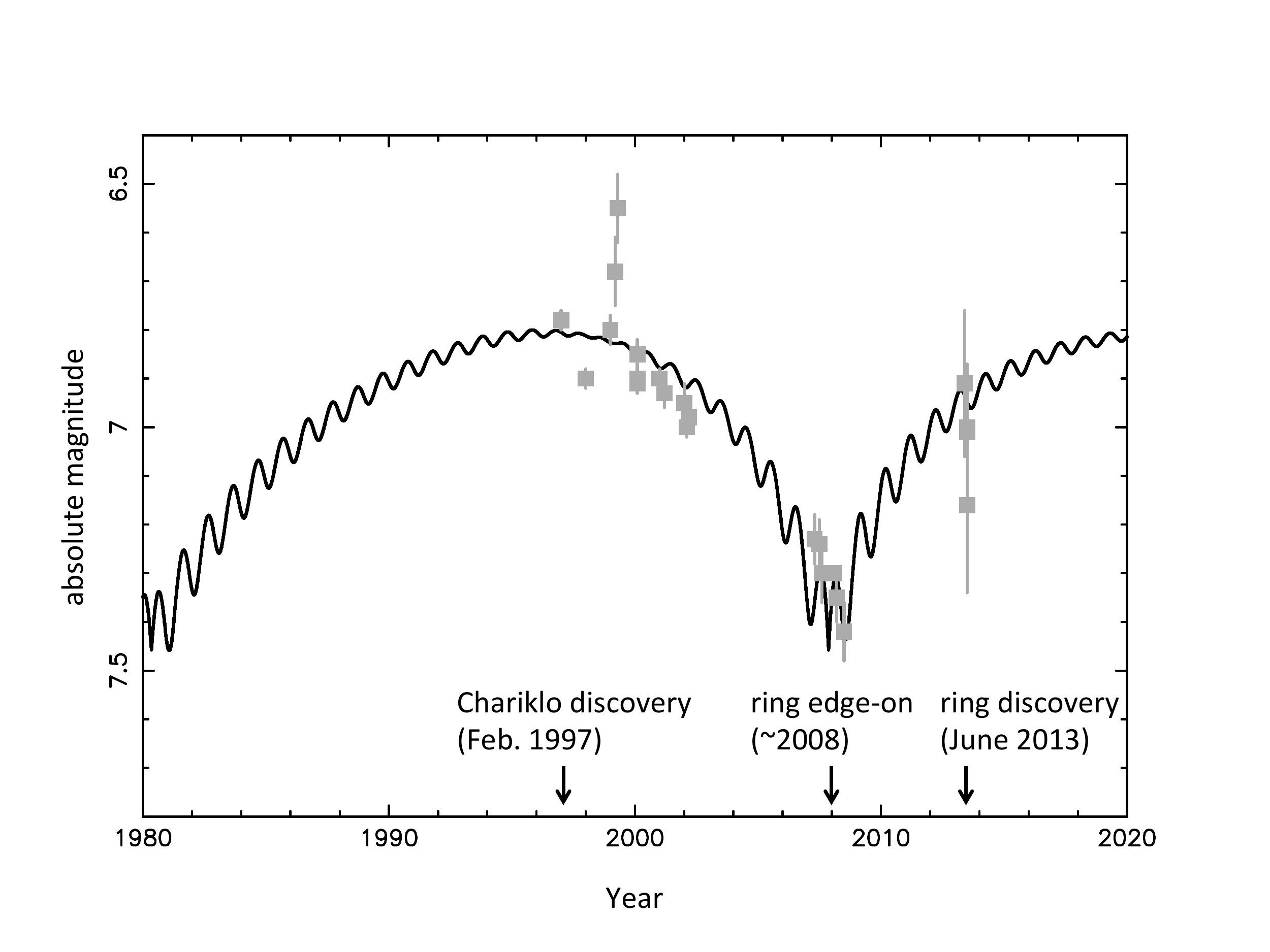}
}
\caption[Chariklo's ring photometry]{%
%
The observed absolute magnitude of Chariklo's system (gray squares)
are fitted by a model (black line) that accounts for both the ring and
Chariklo contributions to the total observed flux (see Eq.~\ref{chap7_eq_abs_mag}).
Adapted from \cite{chap7_duf14b}.
}%
\label{chap7_fig_ring_photom}
\end{figure}

The long-term photometric evolution of Chariklo's system 
is a natural consequence of the changing aspect of its rings. 
During the 63-years orbital period, 
the rings have an opening angle $B$ to the Earth that varies between extreme values 
of about $-60$ to $+60$ degrees.
Using the size, width and radius of Chariklo and its rings
(Tables~\ref{chap7_tab_ck_phys_param} and \ref{chap7_tab_ring_phys_param}),
it follows that the total apparent surface area of the rings at maximum $B$
represents about 35\% of Chariklo's apparent surface area.
For one of the possible ring pole positions previously discussed, the rings 
had their maximum opening angle in 1997 and were observed edge-on in 2008
(Fig.~\ref{chap7_fig_ring_photom}).
The resulting changing apparent ring geometry then  
satisfactorily reproduces the shape and timing of Chariklo's system absolute magnitude,
while excluding the alternative solution aforementioned.

Elaborating on that, let us consider the photometric behavior of a flat, 
circular ring of radius $a$, width $W$ and opening angle $B$.
The brightness of such a flat surface is measured by its reflectivity $I/F$, 
where $\pi F$ is is the incident solar flux density and  $I$ is the intensity 
emitted from the ring surface
(remembering that the reflectivity of a perfect Lambert surface is $I/F=1$).
The flux densities $F_r$ and $F_C$ received from the main ring
and Chariklo are respectively:
\begin{equation}
F_r \propto (I/F)_r S'_r~{\rm and}~F_C \propto  p_C \phi_C(\alpha) S'_C,
\end{equation}
where 
$S'_r = 2\pi a W \mu$ is the ring's apparent surface area, with $\mu=|\sin(B)|$,
$p_C$ is Chariklo's geometric albedo, 
$\alpha$ is the phase angle, 
$\phi_C(\alpha)$ is the phase function (with $\phi_C(0)=1$ by definition), and
$S'_C= \pi R^2_{\rm equiv}$ is Chariklo's apparent surface area, where
the equivalent radius $R_{\rm equiv}$, see Table~\ref{chap7_tab_ck_phys_param}.

Defining $H$ as the absolute magnitude of Chariklo's system (main body plus rings),
i.e. its magnitude at 1 AU from Earth and Sun and at zero phase angle,
and assuming that Chariklo's absolute magnitude $H_C$ is essentially constant over time, 
we obtain:
\begin{equation}
10^{0.4(H_C - H)}= 1 +
\frac{2 a W \mu}{p_C \phi_C(\alpha) R^2_{\rm equiv}}
\left(\frac{I}{F}\right)_r
\label{chap7_eq_abs_mag}
\end{equation}
Monitoring of $H$ vs. time (as $\mu$ changes) 
provides $(I/F)_r$, once Chariklo's photometric properties are known, 
as well as the parameters $a$ and $W$, derived from occultation data
(Table~\ref{chap7_tab_ring_phys_param}).
 
A more detailed modeling of the observed variations 
(\citealt{chap7_duf14b} and Fig.~\ref{chap7_fig_ring_photom}) 
provides 
$(I/F)_r= 0.07  \pm 0.01$ at 0.55~$\mu$m. 
We note that the reflectivity of Saturn's A ring, 
which has an optical depth comparable to that of ring C1R, is 
$(I/F)_{\rm ring~A} \sim 0.3$ \citep{chap7_hed13}.
Conversely, Uranus rings $\alpha$ and $\beta$, 
which also have optical depths comparable to that of C1R, have 
$(I/F)_{\alpha,\beta} \sim 0.05$ \citep{chap7_kar01}.
Thus, Chariklo's rings appear roughly  
three times darker than Saturn's A ring,
twice as bright as Uranus' $\alpha$ and $\beta$ rings, 
and about three times brighter than Chariklo's surface.
Note also that at maximum opening angle ($B \sim 60^\circ$), the 
ring to Chariklo flux ratio is $F_r/F_C \sim$ 0.75. In other words,
the rings significantly contribute to the total flux received from the entire system.

\subsubsection{Composition}
The Chariklo system spectrum has been monitored since 1997, and also shows long-term variations.
In particular, the water ice bands at 1.5~$\mu$m and 2~$\mu$m disappeared
in 2007-08, while being prominent in 1997. The rings provide a natural 
explanation for that behavior: they contain water ice that vanished out of
view during the 2008 ring plane crossing (Fig.~\ref{chap7_fig_ring_photom}).

\begin{figure}
\centerline{
\figurebox{80mm}{}{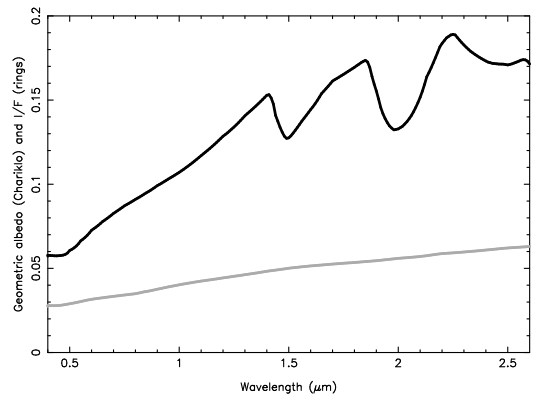}
}
\caption[Chariklo's ring spectrum]{%
Synthetic spectra of Chariklo alone (gray) and its rings (black) 
derived from spectra of Chariklo's system obtained at different epochs, 
with various ring opening angles.
This permits to disentangle the contributions from the main body and from the rings.
Note the water ice bands around 1.5~$\mu$m and 2~$\mu$m in the ring spectrum.
Adapted from \cite{chap7_duf14b}.
}%
\label{chap7_fig_spec}
\end{figure}

By subtracting spectra when the rings are well open and 
spectra of Chariklo alone (edge-on geometry),
one can obtain the spectrum of the rings alone, see Fig.~\ref{chap7_fig_spec}.
It clearly shows the presence of water ice, with a robustly derived abundance close to 20\%
\citep{chap7_duf14b}.
Other compounds must be present, but with far less constrained abundances, 
with degeneracies between the various species.
Current estimates yield values of 
40-70\% silicates, 10-30\% tholins and a small amount of amorphous carbon.
Conversely, Chariklo's spectrum does not reveal any presence of water, and is
consistent with about 60\% of amorphous carbon, 30\% of silicates and 10\% of organics (Ibid.).
Of course, those spectra reveal surface properties, and may be unrelated to
the bulk compositions of the ring particles and Chariklo's interior. It particular, they 
do not preclude the existence of water ice inside Chariklo.


\subsubsection{Central body}

Chariklo has a 
rotation period near 7~hours and an
equivalent radius close to 120~km (Table~\ref{chap7_tab_ck_phys_param}).
Unfortunately, 
other physical properties like size, shape or density are poorly constrained, 
while having important consequences on the ring dynamics, see the next subsection.

Currently, no direct imaging system can resolve Chariklo's disk 
(it subtends less than 0.03~arcsec on the sky), and 
there are not enough stellar occultations observed so far to pin down Chariklo's
size and shape (which can be done in principle at kilometric accuracy
using that method). 
The occultation data currently available show that Chariklo cannot be a spherical body. 
The observations are in fact consistent with either 
an oblate spheroid with equatorial and polar radii of 133$\times$125~km, respectively, or
an ellipsoid with main axes 167$\times$133$\times$86~km, 
with typical uncertainties of 5~km on each dimension 
(Leiva et al. 2016, in preparation).
Assuming a homogeneous body in hydrostatic equilibrium, 
this would imply densities of some
1-3 g cm$^{-3}$ (Maclaurin spheroid case) or close to 
0.8  g cm$^{-3}$ (Jacobi ellipsoid case), with
corresponding dynamical oblatenesses around 0.07 and 0.20, respectively.
Taken together, the two cases considered here imply a Chariklo mass in the range $0.6-3 \times 10^{19}$~kg.

It should be remembered, however, that an irregular shape cannot be currently discarded, 
and that the hydrostatic and homogeneous hypotheses may be invalid.
Clearly, more multi-chord occultations are needed to build a correct model for Chariklo's
size and shape. 

\subsection{Dynamics}
\index{Chariklo's rings, dynamics}

\subsubsection{Roche zone}

Chariklo's tidal disruptive forces must be strong 
enough to prevent the accretion of ring particles into small satellites.
To be disrupted, a particle at distance $a$ from Chariklo should have 
a density $\rho$ of the order of, or lower than a critical density $\rho_{\rm crit}$ 
\citep{chap7_tis13a}:
\begin{equation}
\rho < \rho_{\rm crit}= \frac{4 \pi \rho_C}{\gamma}\left( \frac{R_{\rm equat}}{a} \right)^3,
\label{chap7_eq_roche}
\end{equation}
where $\rho_C$ and $R_{\rm equat}$ are Chariklo's density and equatorial radius, while
$\gamma$ is a dimensionless parameter that describes the structure of the disrupted particles.
For instance,
$\gamma=4\pi/3$ for a sphere (see e.g. \citealt{chap7_mur99}), 
$\gamma \sim$~1.6 for a body that uniformly spreads into its lemon-shaped Roche lobe \citep{chap7_por07}, and 
$\gamma \sim$~0.85  for the (unlikely) incompressible fluid case that corresponds to
the classical Roche limit.
 
All the parameters in Eq.~\ref{chap7_eq_roche} are largely unknown.
The occultation chords obtained in 2013 suggest that Chariklo is 
elongated, with $R_{\rm equat} \sim$~150~km (see above).
%
Moreover,
values of $\gamma \sim$~1.6 seem the most appropriate in the context
of planetary rings \citep{chap7_tis13a}.
Taking $a \sim$~400~km, we obtain $\rho_{\rm crit} \sim 0.4 \rho_C$. 
For an expected icy body like Chariklo, we can assume $\rho_C \sim 1$~g~cm$^{-3}$. 
This suggests that the ring particles should be rather underdense 
($\mathrel{\raise.3ex\hbox{$<$}\mkern-14mu \lower0.6ex\hbox{$\sim$}} 0.5$~g~cm$^{-3}$)
to prevent accretion.
Such densities are actually typical of what is observed in the outer regions of  Saturn's
A ring \citep{chap7_tis13b}.
As previously noted, however, while water ice is clearly identified in Chariklo's rings (Fig.~\ref{chap7_fig_spec}),
other compounds must be present, like silicates or tholins.
This would make Chariklo's ring quite different, in terms of composition, 
from those of Saturn, which are basically pure water ice (see Chapter 3 by Cuzzi \textit{et al.}).
Very little is known about the physical properties of individual ring particles
in general (including those of Saturn's rings).
In that context, it remains to be seen if particles partly composed of
silicates or tholins may have densities as low as 0.5~g~cm$^{-3}$,
for instance if they are  porous or fluffly.
Moreover, the criterion proposed in Eq.~\ref{chap7_eq_roche}  might
miss some of the physics at work and be too crude for a firm claim that
Chariklo's ring particles must be underdense.
%

\subsubsection{Local velocity field and thickness}

Although hugely different in terms of size and mass, Chariklo's
rings share a local velocity field similar to those of Saturn or Uranus.
%
Using a typical mass $M_C \sim 10^{19}$~kg for Chariklo (see above), we obtain
a ring orbital mean motion of  
$n \sim \sqrt{GM_C/a^3} \sim 10^{-4}$ s$^{-1}$ at  $a \sim$~400~km,
where $G$ is the gravitational constant.
This is comparable to the orbital motions in Uranus' rings and the outer part 
of Saturn's rings.
Consequently, the local Keplerian shears, $dv/da= -3n/2$ (where $v$ is the 
orbital velocity), are also comparable.
In other words, a particle in Chariklo's rings esssentially ``sees" the
same local velocity field as a particle in Saturn's and Uranus' rings.

In fact, the mere requirement that the rings must reside inside the Roche zone 
imposes the value of $n$, and thus of $dv/da$.
In effect, combining Eq.~\ref{chap7_eq_roche} and $n= \sqrt{GM_C/a^3}$,  we obtain 
$n \sim \sqrt{\gamma G \rho_{\rm crit}/3}$.
So, the velocity field surrounding the ring particles depends
only on their physical properties, i.e. $\gamma$ and $\rho_{\rm crit}$, 
whatever the central body mass or the ring dimension are.

In the same vein, we see that the ring thickness $h$ only depends on the particle
physical properties, and not on the macroscopic ring parameters.
A dense collisional ring tends to adjust itself so that its Toomre's stability 
parameter $Q$ stays near unity: 
\begin{equation}
Q =  \frac{v_{\rm rms}n}{\pi G \Sigma} \sim \frac{h n^2}{\pi G \Sigma} \sim 1,
\label{chap7_eq_toomre}
\end{equation}
where 
$\Sigma$ is the ring surface density, 
$v_{\rm rms}$ is the ring particle velocity dispersion and
$h$ is the ring thickness, $h \sim v_{\rm rms}/n$.
As Chariklo's main ring is densely packed with particles,  $\Sigma \sim \rho_{\rm crit} R$,
where $R$ is the radius of the largest particles.
So, $h \sim (3\pi)/\gamma Q R \sim$ a few times
$R$ from the estimation of $\gamma$ given before, and from $Q \sim$~1.
In the case of Saturn's rings, $R \sim 1$~m, so that $h \sim 10$~m \citep{chap7_col09}.
We do not know the size distribution in Chariklo's rings, but if it is similar
to that in Saturn's rings, they should also have a thickness of $h \sim 10$~m.

\subsubsection{Mass and angular momentum}

Only rough estimations of the ring mass and angular momentum can be made at the present stage.
%
%
As argued in the previous subsection, the local kinematic conditions in the ring C1R 
should be close to those prevailing in Saturn's rings. 
Assuming a surface density $\Sigma \sim$~500-1000~kg~m$^{-2}$  for C1R
(typical of Saturn's ring densest parts, \citealt{chap7_col09}), 
and considering the quantities in Table~\ref{chap7_tab_ring_phys_param}, 
we obtain a ring mass estimate $M_r \sim 10^{13}$~kg,
equivalent to an icy body of radius $\sim 1$~km.
This corresponds to a very small fraction of Chariklo's mass,
$M_r/M_C \sim 10^{-6}$, and is larger than, 
but still comparable to the corresponding fraction in the case of Saturn's rings, 
$M_r/M_S \sim 10^{-7}$ \citep{chap7_cuz09}. 

Another method can be used to estimate C1R's mass.
Its physical width $W$ varies from about 5.5 to 7.1~km 
(Table~\ref{chap7_tab_ring_phys_param}). 
This suggests that C1R may behave like some of the Uranian rings \citep{chap7_fre91}, i.e. 
a set of nested elliptical streamlines locked into a common precession rate regime,  
against the differential precession (stemming from the central body's oblateness)
that should destroy this configuration.

In those models, the narrow ring is globally described by
an ellipse with mean semi-major axis $a$ and mean eccentricity $e$,
while its inner and outer edges are described by aligned ellipses with 
semi-major axes $a_{\rm inn}$ and $a_{\rm out}$, and 
eccentricities $e_{\rm inn}$ and $e_{\rm out}$, respectively. 
To first order in $e$, the width of the ring then varies with true anomaly $f$  
as $W= \Delta a [1-q_e\cos(f)]$,
where $q_e= e + a \partial e/\partial a$ is a dimensionless parameter that depends 
on both the eccentricity and its gradient across the ring, 
$\partial e/\partial a= \Delta e/\Delta a$, with
$\Delta e= e_{\rm out}- e_{\rm inn}$ and 
$\Delta a= a_{\rm out}- a_{\rm inn}$.

One mechanism proposed by \cite{chap7_gol79a,chap7_gol79b} to lock the streamlines 
into a rigid precession regime is self-gravity. In essence, the mass of the inner half of the ring
increases the precession rate of the outer half, and vice-versa, thus maintaining the alignment.
This requires a ratio $M_r/M_C$ of the order of 
$(e/\Delta e) (\Delta a/a)^3 J_2 (R_C/a)^2$, 
where $J_2$ is the dynamical oblateness of the central body.
Elaborating on that basis, \cite{chap7_pan16} estimate a C1R mass of a few $10^{13}$~kg,
comparable to the estimate already given before.
This reinforces the notion that the ring C1R is comparable, both in terms of surface density and 
dynamical behavior, to some of the dense and narrow rings of Saturn or Uranus.

Those estimates must be considered with caution, though, 
first because neither $e$ nor $q_e$ are currently known.
The occultation data are not yet accurate and numerous enough to provide detailed 
ring orbital solutions and edge models.

Secondly, 
only variations of width $W$ with respect to the \textit{inertial} mean longitude $\lambda$ have been derived right now, 
while variations vs. true anomaly $f$ should be determined to test rigid precession models.
This will be possibe only when the ring apsidal precession rate 
$\dot{\varpi} \sim 1.5 (R_c/a)^2 J_2 n$ is determined.
An expected value of $J_2$ is $\sim \Omega_C^2 R_C^3/2GM_C$, 
assuming a homogeneous body, where $\Omega_C=2\pi/P_C$ is Chariklo's spin rate. 
From Table~\ref{chap7_tab_ck_phys_param}, 
one obtains $J_2 \sim 0.08$ and
$\dot{\varpi} \sim 10^{-6}$~s$^{-1}$, so that the ring apse should precess 
over a period of a couple of months only. 
This is much shorter than the eleven months or so separating the June 2013 and April 2014 occultations 
from which values of $W$ are derived (Table~\ref{chap7_tab_ring_phys_param}).
Consequently, it is not yet possible to compare  consistently those observations
and obtain a coherent plot of $W$ vs. $f$.

Finally, we do not know yet if the observed width variation is caused by a $m$=1
azimuthal wavenumber, or by some higher (free or forced) wavenumbers 
which would require a revision of the mass estimation made above.

More generally, the physics at work in dense narrow rings may be more complex 
than the purely self-gravitating models evoked so far.
In particular, viscous effects due to interparticle collisions near sharp ring edges  
resonantly perturbed by (yet to be discovered) shepherd satellites may significantly increase the 
mass estimation quoted above, see \cite{chap7_chi00} and \cite{chap7_mos02}.
Those models predict enhanced ring surface densities at some 100-500~m from the edges, 
consistent with the double-dip structures observed in $\alpha$ and $\epsilon$ Uranus' rings occultation 
profiles \citep{chap7_fre91}, 
and interestingly,  in the C1R profile too (Fig.~\ref{chap7_fig_ring_prof}).
 
Finally, approximating Chariklo as a homogeneous sphere of radius $R_{\rm equiv}$, 
the ratio of the ring angular momentum to that of Chariklo is
$H_r/H_C \sim (M_r/M_C) P_C \sqrt{G \rho_C} \sqrt{a/R_{\rm equiv}}$.
From Tables~\ref{chap7_tab_ck_phys_param} and \ref{chap7_tab_ring_phys_param}, 
we obtain $H_r/H_C \sim 10^{-5}$.
Applying the same calculation to Saturn's rings, where we  assume that the rings 
are uniformly spread between $\sim$~92,000 and 137,000~km, we obtain 
a ratio $H_r/H_S \sim 10^{-6}$ (using $M_r/M_S \sim 10^{-7}$, see above). 
This is smaller than, but still comparable to the fraction $H_r/H_C$.
%
%
In any case, we see that
very small fractions of Chariklo's mass and angular momentum are stored
in the rings, a noteworthy result when it comes to discussing the rings' origin.

\subsubsection{Putative shepherd satellites}

Rings C1R and C2R are both sharply confined, see Fig.~\ref{chap7_fig_ring_prof}.
If unperturbed, they should spread out on a timescale of \citep{chap7_gol79a}:
\begin{equation}
t_\nu \sim \frac{W^2}{\nu},
\label{chap7_eq_spread}
\end{equation}
where 
$\nu$ is the kinematic viscosity associated with particle collisions.
A typical value of $\nu$ is $\sim n h^2$, where $h$ is the ring thickness.
Taking $W \sim 5$~km (Table~\ref{chap7_tab_ring_phys_param}), 
we obtain  $t_\nu \sim 10^4/h^2$~years, where $h$ is expressed in meters.
Assuming again $h \sim 10$~m, 
we obtain $t_\nu \sim$ a few thousand years. 
Moreover, Poynting-Robertson (PR) differential drag also causes
a spreading over a timescale of \citep{chap7_gol79a}:
\begin{equation}
t_{PR} \sim \left(\frac{c^{2}}{4 f_\odot}\right) \left(\frac{W}{a}\right) \rho \tau R,
\label{chap7_eq_PR}
\end{equation}
where 
$c$ is the velocity of light, 
$f_\odot$ is the solar flux density at Chariklo,
$\rho$ the density of the particles and
$\tau$ is the optical depth.
Taking $\rho < \rho_{\rm crit} \sim 0.5$ g~cm$^{-3}$, as explained before, 
and $\tau \sim 1$, we obtain a typical value of $t_{PR} \sim 10^9 R_{\rm meters}$~years, 
i.e. a few million years for sub-cm particles.
The effect of PR drag is even more drastic for smaller grains,
with a depletion time of only a few months  for micrometric particles. 
Even if very crude, these estimations show that 
Chariklo's rings are either very young, or confined by an active mechanism.

It is remarkable that apparently similar ring confinement occurs in systems 
so widely different (in terms of orbital radii) as those of Saturn or Uranus, 
compared to Chariklo. 
In fact, looking at Chariklo's rings' optical depth profiles (Fig.~\ref{chap7_fig_ring_prof}), 
it is hard to distinguish them from their Uranian cousins, 
see \cite{chap7_fre91} and Chapter~4.

A classical theory to confine ring material invokes the presence of putative ``shepherd satellites".
In its simplest version, a shepherd of mass $M_s$ can exert a torque $T_m$ 
onto a ring edge at a discrete $m+1:m$ mean motion resonance 
(where the ring particle completes $m+1$ revolutions while the satellite completes $m$ revolutions, 
with $m$ integer), see \cite{chap7_gol82}:
\begin{equation}
T_m \sim 8.5m^2 a^4 n^2 \Sigma \left(\frac{M_s}{M_C}\right)^2.
\label{chap7_eq_torque_m}
\end{equation}
As $m$ increases, resonances overlap and the torque density 
(torque per unit interval of semi-major axis) is:
\begin{equation}
\frac{dT}{da} \sim 2.5 a^3 n^2 \Sigma \left(\frac{M_s}{M_C}\right)^2 \left(\frac{a}{x}\right)^4,
\label{chap7_eq_torque_over}
\end{equation}
where $x$ is the distance between the satellite and the ring.
To prevent spreading, the satellite torque must balance the viscous torque $T_\nu$
associated with inter-particle collisions. In a Keplerian velocity field, it is:
\begin{equation}
T_\nu = 3\pi n a^2 \nu \Sigma.
\label{chap7_eq_torque_nu}
\end{equation}
Making $T_m = T_\nu$, and in the case of discrete resonances, 
the radius $R_s$ of a shepherd  with density $\rho_s$ is: 
\begin{equation}
R_s
\sim \left(\frac{\rho_C}{\rho_s}\right)^{1/3}  \left(\frac{h}{m a}\right)^{1/3} R_C.
\label{chap7_eq_shepherd_size}
\end{equation}
This yields radii of a few km for icy shepherds ($\rho_s \sim 1$~g~cm$^{-3}$), 
taking $m$ of a few times unity and $h$ a few meters.
Concerning the gap between C1R and C2R, it should be opened in the overlapping resonance
regime (Eq.~\ref{chap7_eq_torque_over}), in which case, the radius of the satellite is 
\citep{chap7_gol82}:
\begin{equation}
\frac{R_s}{R_C} 
\sim \left(\frac{\rho_C}{\rho_s}\right)^{1/3}  \left(\frac{h}{a}\right)^{1/3}  \left(\frac{W_{\rm gap}}{a}\right)^{1/2}
\mathrel{\raise.3ex\hbox{$<$}\mkern-14mu \lower0.6ex\hbox{$\sim$}} 1~{\rm km}
\label{chap7_eq_gap_opener}
\end{equation}
where $W_{\rm gap} \sim$ 8.5~km is the full width of the gap
(Table~\ref{chap7_tab_ring_phys_param}).
Note in passing that the mass of the shepherds estimated here would be  comparable to that of the  
rings (see the previous subsection). In other words, there would be roughly the same amount
of material in the form of rings and in the form of (putative) shepherd satellites.

This said, this model poses new problems. 
In effect, as the shepherd confines a ring through a torque $T_m$,
the reaction from the latter induces a migration rate $|\dot{a}_s| \sim 2|T_m|/(anM_s)$, 
where $a_s$ is the shepherd's semi-major axis.
Considering that $T_m = T_\nu$ and using Toomre's criterion of Eq.~\ref{chap7_eq_toomre},
one obtains:
\begin{equation}
|\dot{a}_s| \sim 
36m \left(\frac{h}{a} \right)  \left(\frac{h}{T_{\rm orb}}\right),
\label{chap7_eq_recess}
\end{equation}
where $T_{\rm orb}$ is the orbital period.
We may apply this formula to the shepherd satellites Cordelia and Ophelia
which confine the Uranian $\epsilon$ ring. In this case, 
$m \sim 10$,
$T_{\rm orb} \sim 10$~hours, 
$a \sim 50,000$~km and  
$h \sim 10$~meters. 
This yields $|\dot{a}_s| \sim 1$~m~s$^{-1}$. As the shepherds orbit at some
1000-2000~km from the $\epsilon$ ring, this implies short recession timescales 
of some million years for those two satellites.
This problem is exacerbated for Chariklo, due to the smallness of the semi-major axis $a$ 
(by more than two orders of magnitudes) compared to the case of the giant planets.
Applying Eq.~\ref{chap7_eq_recess} to the Chariklo case actually provides recession timescales
of some thousands of years only if one assumes again $h \sim 10$~meters, and considering that
$T_{\rm orb}$ must again be of the order of 10~hours.

One possibility is that, for a so far unexplained reason, Chariklo's ring particles
are much smaller than those of Saturn or Uranus, resulting in a much smaller value
of $h$, and thus, much longer recession timescales for the shepherds.
The shepherding physics may also be more complex than assumed here.
For instance, the viscous torque~(\ref{chap7_eq_torque_nu}) 
can be significantly reduced due to the local reversal of the viscous
angular momentum flux, caused by the satellite itself \citep{chap7_gol87}. 
This in turn would reduce the masses of the shepherds estimated above,
as well as their migration rates.

In summary, and except if Chariklo's rings are very young, 
some deep understanding of the shepherding mechanism, 
and in particular a better knowledge of the ring local collisional dynamics 
are required to better assess the short timescale problems described above.
In that context, detections of the putative shepherd satellites would be very 
helpful to understand Chariklo's ring confinement,
but this remains a very challenging observational task.

At this point, it is worth mentioning that resonances may arise not from satellites,
but from the very shape of Chariklo.
For instance, a topographic feature of some 5~km in height on Chariklo's surface
might cause (tesseral-type) 
resonant perturbations that are comparable in strength to those stemming from 
the putative satellites mentioned above. 
The same is true if the body is elongated in one direction by a few kilometers, 
in which case non-axisymmetric 
perturbations arise from the bulges associated with the elongation.
Assuming thet Chariklo's mass is in the range $0.6-3 \times 10^{19}$~kg,  
the corotation radius 
-- where particles have an orbital period matching that of Chariklo ($\sim$~7 h) --
would lie somewhere between 185 and 320~km. 
On each side of this corotation radius, 
first order commensurabilities $m+1:m$  between the particle mean motion
and Chariklo's orbital period would appear. 
It is instructive to note that the 2/1 outer resonance 
(corresponding to particles with orbital period close to 14~h) should occur somewhere 
between 290 and 510~km, bracketing the region where the rings are found.
It is too early to conclude anything before accurate 
measurements of Chariklo's mass and shape are made, 
but it is worth remembering that the ring dynamics might
be significantly influenced by resonances with the spin of the central body.

\subsection{The origin of Chariklo's rings}
\index{Chariklo's rings, origin}

The general portrait that emerges from the previous subsections 
is that of a ring system composed of underdense particles
($\rho \mathrel{\raise.3ex\hbox{$<$}\mkern-14mu \lower0.6ex\hbox{$\sim$}} 0.5$~g~cm$^{-3}$),
partially composed of water ice, and
confined by small shepherd satellites of some kilometers in size
that contain a mass comparable to that of the rings.
Moreover, a very small fraction of mass ($\sim 10^{-6}$) and 
angular momentum ($\sim 10^{-5}$), compared to Chariklo, 
are required to explain the observed rings.

\subsubsection{Rings around other small bodies}

At present, it remains unclear whether Chariklo's rings
are generic and frequent features around small bodies, 
or are an exceptional system resulting from a fine tuning between
various physical properties.
Hundreds of Main Belt asteroid occultations have been monitored, 
but no report of secondary events possibly due to rings have been reported so far. 
A handful of occultation events involving TNO's have been published
up to now \citep{chap7_ell10,chap7_sic11,chap7_ort12,chap7_bra13}, 
and again no evidence of ring events have been documented.
It should be noted, however, that Chariklo's rings cause very brief 
stellar drops (at sub-second level, see Fig.~\ref{chap7_fig_danish})
that are easily overlooked if integration times and/or noise levels are too large. 
Also, re-analysis of the best occultation data sets obtained so far might
reveal ring-related features.
Moreover, imaging such systems is challenging from Earth.
For instance, Chariklo's rings do not span more than 0.04 arcsec around the main body. 
This makes direct detection very hard, even on the best instruments available nowadays.
So, other ring systems may still be undiscovered due
to the lack of high-quality occultation observations or high resolution imagers.

\subsubsection{The case of Chiron}

The object (2060) Chiron is the second largest Centaur known to date, 
with a diameter of  $218 \pm 20$~km \citep{chap7_for13} and 
perihelion-aphelion distances of 8.4-18.8 AU.
Two stellar occultations observed in 1993 and 1994 
actually revealed secondary events that were interpreted as due to collimated cometary jets
\citep{chap7_ell95,chap7_bus96}.
A more recent Chiron occultation in 2011 revealed symmetric, narrow and sharp 
double dips, similar in depth and width to those observed around Chariklo.
They have been interpreted as being due to a spherical shell surrounding Chiron \citep{chap7_rup15}
or to a ring system akin to those of Chariklo.

The ring hypothesis is supported by various arguments \citep{chap7_ort15}:  
(1) the strong similarity between the event reported by \cite{chap7_rup15} 
and the one in Fig.~\ref{chap7_fig_danish}, 
(2) the fact that the reconstructed ring orientation globally explains the  photometric behavior of Chiron since 
$\sim$~1980, with a minimum in 2001 when the rings should be observed edge-on, 
(3) the fact that this also explains the variations over time of Chiron's rotational lightcurve amplitude, 
assuming that the rings lie in the equatorial plane of a triaxial central body, and
(4) the fact that Chiron's spectrum exhibits variations of the water ice band, 
with a disappearance in 2001 that would  be caused by an edge-on geometry of the ring
at that epoch, as it is the case for Chariklo in 2008.
However, the ring hypothesis remains debatable as Chiron has cometary activity 
\citep{chap7_mee89,chap7_luu90} that may skew the interpretation. 
Moreover, the 1994 Chiron occultation showed one sharp secondary event followed
by a more diffuse stellar drop that is incompatible with azimuthally uniform narrow and 
dense rings, perhaps calling for the existence of incomplete rings (arcs) instead.
Discriminating between the shell and ring interpretations (or others) clearly requires higher-quality 
multi-chords occultations.

In any case, it is interesting to see that Chiron, 
a Centaur similar to Chariklo in terms of size and orbit, 
is also surrounded by optically dense material.
This suggests that common physical processes may be at work, 
giving rise to unaccreted closeby material.

\subsubsection{Are Centaurs special?}

Several circumstances can make Chariklo (and possibly other Centaurs) 
special to allow it to possess rings: 
$(i)$ its heliocentric distance, 
$(ii)$ possible transient cometary activity,
$(iii)$ its size, and
$(iv)$ significant gravitational perturbations from the giant planets.

Point $(i)$ is discussed by \cite{chap7_hed15}, who considers that 
water ice particles may have the adequate physical properties 
to meet the condition given in Eq.~\ref{chap7_eq_roche} between 8 and 20 AU.
More precisely, in that heliocentric range, water ice may reach a typical temperature of 70~K,  
low enough for the ice to avoid sublimation 
(which explains why no rings are seen around Main Belt asteroids), 
but still high enough to remain weak and consequently be subjected to tidal disruption in the Roche zone.
This would be consistent with the fact that Chariklo travels
between 13 and 19 AU from the Sun.

Another circumstance linked to heliocentric distance is the
lower impact velocities prevailing in Chariklo's region 
- typically 1~km~s$^{-1}$ -
compared to relative velocities in the Main Belt,  
about 5~km~s$^{-1}$.
This results in much less destructive collisions that may give rise to a debris disk
around the impacted body, from which rings form, 
instead of dispersing 
most of the pieces to infinity.

Turning to point $(ii)$, it appears that some Centaurs have cometary activity.
This is documented for Chiron \citep{chap7_mee89,chap7_luu90} and Echeclus \citep{chap7_rou08},
a Centaur with perihelion-aphelion distances of 5.8-15.6 AU and diameter of about 65~km
\citep{chap7_duf14a}. %
In Echeclus' case, the cometary-like episode of March 2006 
showed a coma around a source that was \textit{separated} from Echeclus.
This coma may have been caused by a 8-km object moving near the main body,
that could be a satellite or a fragment ejected from the surface.
This adds one case in the list  of Centaurs surrounded by (in this case, transient) material.

The escape velocity at the surface of a body of density $\rho$ and radius $R$ is 
\begin{equation}
v_{\rm esc}= \sqrt{\frac{8\pi G \rho}{3}}R \sim 0.75 R_{\rm km} {\rm ~m~s}^{-1},
\end{equation}
assuming an icy body ($\rho \sim 1$~g~cm$^{-3}$).
This implies a rough value of $v_{\rm esc} \sim 100$~m~s$^{-1}$ for Chariklo or Chiron. 
This is typical of the upper limit for terminal velocities of dust
grains in a cometary coma \citep{chap7_del82,chap7_ten11}.
This coincidence raises the interesting possibility that Chariklo's rings
have an endogenous origin. They could be formed of cometary material
ejected from the surface with velocities large enough to prevent 
an immediate in-fall onto the surface, as is the case for Triton's geyser material, 
but still  small enough to prevent escape to infinity, as is the case for km-sized comets.

In any case, no dust production has been observed so far for Chariklo \citep{chap7_gui09}, 
with an upper limit of about 2.5~kg~s$^{-1}$ for the dust production rate \citep{chap7_for14}.
Similarly, no gas production has been found for that body, 
with typical upper limits of 
$2 \times 10^{28}$ molecules~s$^{-1}$ of CO ($\sim 10^3$ kg s$^{-1}$) and 
$8 \times 10^{27}$ molecules~s$^{-1}$ of HCN ($\sim 400$ kg s$^{-1}$), see
\cite{chap7_boc01}.
In that context, deeper searches for dust or gas production 
might reveal a low level cometary activity for Chariklo, 
and thus constrain ring origin models.


Finally, we already noted that 
Chariklo is moving on an unstable, short life-time (10 Myr) orbit controlled by 
Uranus \citep{chap7_hor04}. The encounter distance $\Delta_{\rm disrupt}$ 
at which a ring of semi-major axis $a$ is disrupted is given by the Hill sphere radius
$\Delta_{\rm disrupt} \sim a (3M_U/M_C)^{1/3}$, where $M_U$ is the Uranus mass.
Using 
$M_U \sim 10^{26}$ kg, 
$M_C \sim 10^{19}$ kg and $a \sim 400$~km
yields $\Delta_{\rm disrupt} \sim 5$ Uranian radii. 
Current estimations show that such a small distance has a small probability to occur
(some 10\%, see \citealt{chap7_hyo16}) during Chariklo's migration,
and that globally Chariklo's ring  system can survive more than 90\% of its encounters
with giant planets \citep{chap7_ara16}.
In other words, if Chariklo's rings were already formed while the object was in the 
TNO region, they should have safely experienced the migration episode.

However, encounters with Uranus may have destabilized 
a pre-existing, marginally stable Chariklo's \textit{satellite} system, 
causing orbit crossing and then collisions between the satellites,
resulting in ring formation, a scenario that is still to be investigated.
Note finally that a mere disruption of Chariklo during a close 
encounter with Uranus  may also be envisaged.
This might result in a debris disk around the body, from which 
small moons shepherding ring material could emerge 
near Chariklo's Roche limit \citep{chap7_hyo16}.

\section{Rings around natural satellites}
\index{Satellites, rings around}

Currently,
no rings have been confirmed orbiting natural satellites in our solar system, 
but physical evidence suggests that such rings could have been a common occurrence in the past.  
In fact, equatorial features occur on two moons of Saturn, Rhea and Iapetus, 
and could be remnants of such rings, as discussed below.

\subsection{Iapetus}

\begin{figure}
\centerline{
\figurebox{80mm}{}{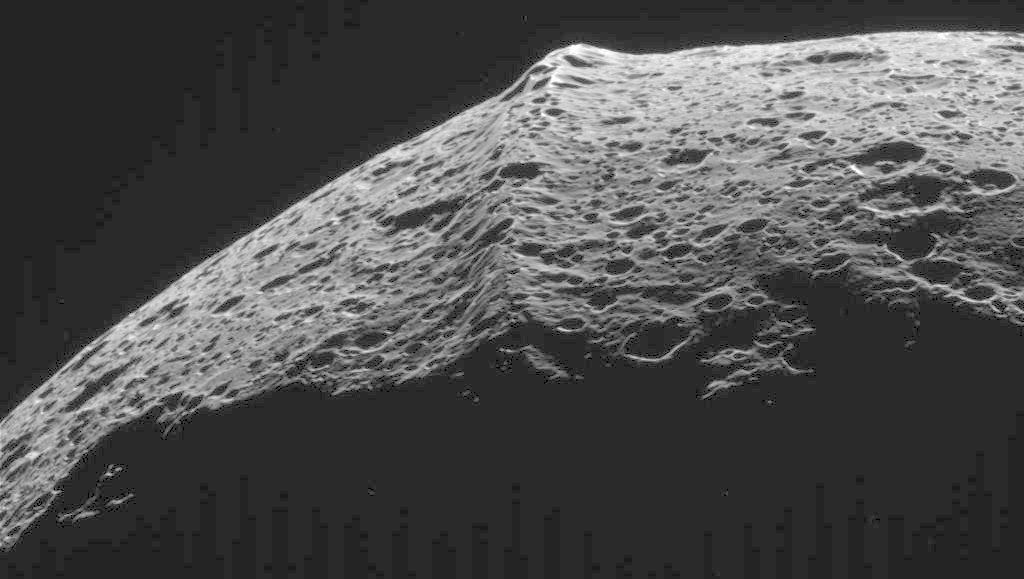}
}
\caption[Iapetus equatorial ridge]{%
An image of the ridge of Iapetus taken from around 62,000 km in September 2007 
by the Cassini-Huygens mission 
(image number N00091828: credit to NASA/JPL-Caltech/Space Science Institute).
}%
\label{chap7_fig_iapetus_ridge}
\end{figure}

The ridge on Iapetus is a puzzling feature, 
up to 20 km tall and about 100~km wide,
see \cite{chap7_den05a,chap7_den05b} and Fig.~\ref{chap7_fig_iapetus_ridge}.
%
It is challenging to reconcile this with the global history of Iapetus.
One approach has modeled it as the outcome of endogenic (tectonic or other) processes, 
but none could satisfy the observational constraint of a single equatorial high-standing ridge, 
see \cite{chap7_cas07} and \cite{chap7_rob10}.
Alternately, an exogenous origin has been proposed - the
remnants of a ring that has fallen to the surface
(\citealt{chap7_ip06,chap7_lev11,chap7_dom12}).
Following this latter idea, one can use the constraining observed features of the
Iapetan ridge - its dimensions, morphology/slopes and some
localised cases of parallel ridges or tracks - and combine them with
the dynamics of the Saturnian system and a ring's tidal evolution to
estimate the origin and properties that this proposed ring could have
had.

Note that the Voyager imaging had not high enough resolution at equatorial
latitudes to detect directly any significant equatorial topography 
\citep{chap7_smi81,chap7_smi82}. 
However, the analysis of the limb data many years
later strongly pointed to a massive mountain-like structure with
heights up to 20 km 
\citep{chap7_den00}.
These were identified in the Cassini Regio (the dark leading side) between 
the longitudes of 180 W and 200 W.

The most significant imaging of the equatorial ridge was done by the
Cassini spacecraft during a flyby on 31 December 2004. The dark,
leading side, was reported by 
\cite{chap7_por05} 
to have an equatorial ridge up to 20 km in height, confirming Denk's finding. 
Profiles from the Digital Terrain Model (DTM) constructed from Cassini data find a diversity in the morphology
of the ridge. 
At times it has a strictly steep-sided triangular shape
(see ridge profiles r1 and r2 in Fig. 5 of \citealt{chap7_gie08}),
while at others a more ``flat-top'' or trapezoidal shape characterized by lower slopes on the
Northern side estimated to decrease to 15 deg down to 8~deg 
in ridge profile r3 and 4~deg in ridge profile r4.  
Subsequent global mapping of the topography \citep[by PMS, published in][]{chap7_dom12} shows that the ridge is discontinuous with a series of linear ridges and isolated quasi-conical massifs along the equatorial trace.  The elevation is variable along the length and appears to be absent in some stretches, though it is essentially global in character.

The trailing side DTM was constructed of only two sets
of stereo pairs, but reveals that the ridge continues on the back
hemisphere, and appears to be centered at 30 deg East 
\citep{chap7_gie08}.
Given the current observations, with confirmed existence of
the ridge spanning the entire well-observed leading side hemisphere
\citep{chap7_por05,chap7_gie08}, 
the leading side limb and also
the trailing side 
\citep{chap7_gie08}
it is safe to assume that this feature is indeed global.

The size and the equatorial location of the ridge made it an enigmatic
feature, but the larger context at Iapetus also includes its
synchronous spin state, and its non-equilibrium global shape. Iapetus
is synchronously locked with its orbital period around Saturn of 79
days, and estimates based on solid-body tides suggest that de-spinning
to this state would take $>$~10 Gyr 
\citep{chap7_pea77}.
Meanwhile, the global shape of Iapetus suggests it is not in
hydrostatic equilibrium, 
as it has a shape expected for a body with a 16-h rotation rate 
\citep{chap7_tho07,chap7_cas07,chap7_tho10}.
Combined, the features of the equatorial ridge and the spin state and global
shape of the entire icy satellite point to a complicated history, one
in which the ridge could have formed from an in-falling ring 
\citep{chap7_ip06}.


Two distinct takes on this idea have been explored. 
\cite{chap7_lev11} 
propose that Iapetus suffered a violent impact, and ejecta
formed a large disk of debris that quickly damped to the equatorial
plane straddling the synchronous limit for a then-faster spinning
Iapetus (as the global shape implies a faster rotation in its past) --
similar to the proto-lunar disk at Earth. A sub-satellite forms from
the disk beyond the Roche limit and the synchronous limit, and its
tidal interactions with the remaining debris push the ring to the
surface of Iapetus forming the equatorial ridge. The sub-satellite
then tidally evolves away from Iapetus, slowing the rotation of
Iapetus and aiding its de-spinning until it is eventually stripped
from the orbit of Iapetus.

\cite{chap7_dom12}
propose a similar scenario, but with a different
origin and fate for the formed sub-satellite. Here, a $\sim$~100~km
body impacts Iapetus and is captured into orbit -- similar to the
Pluto-Charon formation scenario. If the orbit is retrograde it will
tidally evolve inward until reaching the Roche limit, after which it
will be tidally disrupted and eventually rain down to the surface
piece by piece building the equatorial ridge.

While there are some differences in the models, they both propose the
existence of a substantial ring orbiting Iapetus in its
past. Fundamentally, both models would require a ring with a minimum
mass equal to that needed to create the equatorial ridge. Estimates
for this mass depend on the assumed shape and size of the ridge before
its degradation due to bombardment over Solar System timescales - it
has been estimated between $5.5 \times 10^{17}$ and $4.4 \times 10^{19}$ kg 
\citep{chap7_ip06,chap7_lev11},
or a fraction up to a few percent of the mass of Iapetus. 
The 
\cite{chap7_lev11}
model would actually require more mass in the
ring initially, as it would eventually build a sizable sub-satellite
leaving behind enough mass in the ring to then fall to the surface
and build the ridge.

Neither model predicts a long lifetime for the ring. 
\cite{chap7_lev11}
find a spreading timescale of only 100's of years owing to
the strong tidal interaction with the sub-satellite. Meanwhile, the
timescale for the survival of a ring is longer in 
\cite{chap7_dom12}
as there is no external interaction with other orbiting debris - the
sub-satellite is tidally disrupted and then falls piecewise onto the
surface on tidal timescales. This was estimated to be quite rapid
compared to the 10-100's~Myr tidal timescales of the system, owing to
the eventual ring having a very high surface mass density and
therefore rapid viscous evolution. The timing of the formation,
evolution and demise of the ring is different in each case as
well. The ring in 
\cite{chap7_lev11}
would form and be lost coincident
with the large impact and birth of the sub-satellite, with a longer
tidal scenario to play out with the sub-satellite. The 
\cite{chap7_dom12}
scenario would have the ring form  well after the impact
and tidal evolution of the sub-satellite, so at the end of the
story. While these may not be testable distinctions for Iapetus with the
current dataset, they do provide scenarios that could be distinguished
in other parts of our Solar System or external to our Solar System.

While these are both compelling models, much of their viability lies
in the geophysical evolution of Iapetus, the large uncertainties in
tidal evolution timescales, and even in the impact physics of ring
particles pummeling the surface. Further advances could be made by
better understanding the stratigraphy on the surface of Iapetus to
date the ring and understand its relationship to some of the largest
impacts. 

\begin{figure}
\centerline{
\figurebox{80mm}{}{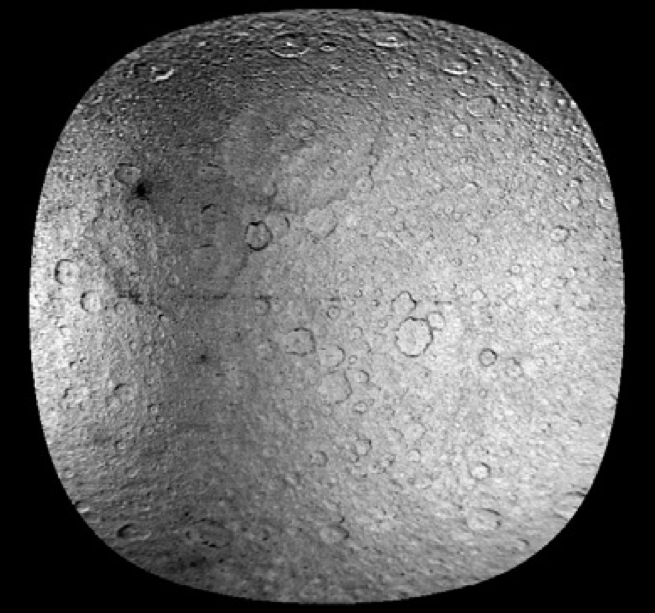}
}
\caption[Rhea ``blue pearls'']{%
This view shows a Cassini color ratio map (IR3/UV3) of the anti-Saturnian facing hemisphere of Rhea.  
Such maps reveal discrete spots along the equator 20-50~km across that have unusual spectral signatures 
enhanced in the ultraviolet, giving them a ``bluish" appearance in Cassini 3-color mosaics.
Each spot correlates with local highs in topography.  
Effective resolution is $\sim$1.75 km/pixel.
}%
\label{chap7_fig_rhea_blue_pearls}
\end{figure}

\subsection{Rhea}

In 2009 a putative ring was reported in orbit around Rhea, 
after sharp drops in measured electron counts from the fields and particles instruments, symmetric about the moon, were detected by the charged-particle detectors 
aboard the Cassini spacecraft \citep{chap7_jon08}.
However, deep searches in both high-phase and low-phase Cassini ISS narrow-angle camera images ruled out dust material orbiting Rhea  as an explanation for those drops
\citep{chap7_tis10}. 
After this report, independent workers examining multispectral ISS mosaics of Rhea discovered ``blue pearls" on the surface of that moon (\citealt{chap7_sch11}).
%
These features take the form of irregularly shaped surface patches 5-50 km wide and 50-150 
km apart, visible only as color anomalies, and only along the equator of Rhea (Fig.~\ref{chap7_fig_rhea_blue_pearls}). %
These patches are brighter in the near-UV and hence have a bluish signature when mapped 
in UV-green-IR filter combinations.  The key observation at Rhea is that these spots or ``Blue Pearls" 
form only on the highest local topography along the equator 
\citep{chap7_sch11}, 
consistent with infalling orbiting debris striking the surface in the manner suggested by 
\cite{chap7_ip06} for Iapetus.  No constructional mountains or ridges are associated with these features, however, as the relief along the equator consists of pre-existing crater rims and ridges. 
The \cite{chap7_sch11} proposal was that fine material spiraled in toward the surface and disturbed regolith 
on high topographic ridges.  Due to the very low impact angles, lower lying topography ``downstream"
of these impact points would be shielded from impact and thus uncolored. 

\subsection{Comparison of Iapetus and Rhea features}

By itself, the equatorial ridge on Iapetus remained enigmatic, though all efforts to identify a tectonic mechanism failed.  
The discovery of a second non-tectonic equatorial feature on Rhea strengthens the case for ring deposition. 
The major difference at Rhea (compared to Iapetus) is that there is no significant accumulation 
of a topographic deposit at Rhea, suggesting that the mass of the ring system there was also much less than at Iapetus.   

This said, two major features link the Rhean Blue Pearls and the Iapetus ridge.  
The first is the narrow equatorial great circle pattern of both.  The second is the discontinuous nature of both. 
The Blue Pearls are widely separated and occur only on high-standing topography.  
The Iapetus ridge is also discontinuous and forms widely separated sub-ridges and promontories along the equator 
(Fig.~\ref{chap7_fig_iapetus_ridge}s). 
In the case of a low-mass ring system, only the tops of high-standing features crossing the equator 
would be disturbed by infalling debris (as at Rhea).  
In the case of a massive ring system, material would accrete ``backwards" along the equator at 
high standing blocking ridges, forming a partial ridge.  
With enough mass, a more continuous ridge is built up.

The two moon putative ring deposits appear to be of different ages.  
The Rhea color anomaly is a surficial effect and easily erased.  
The inference is that it is not ancient as blue color signatures fade over time and are not visible in older eroded craters.  
The small-scale complexity of these spots also points to a relatively young age 
as small scale features always tend to be ground up into the regolith of airless planetary bodies within a billion years or so, 
depending on location.  
The Iapetus ridge is much more massive and not easily destroyed or eroded.  
Landslides have affected its profile \citep{chap7_sin13} 
and it is heavily cratered.  
Some large craters cut into it.  It is inferred that it is very ancient, but also it does not appear 
to be primordial as its profile would be much more dissected than it appears.  
We infer that it formed sometime within the first 1-2 Gyr of formation but that tighter constraints on age are lacking.

Finally, 
no equatorial features (ridge, color anomaly, or otherwise) have been observed on other icy moons of Saturn, while mapping on the Uranian satellites is limited by poor coverage of the equatorial regions. %
Color and topographic mapping are at least as good on the other Saturnian satellites as they are on Rhea and Iapetus, 
and if such ring deposits ever formed they have either been erased by subsequent bombardment or E-ring deposition \citep[e.g.,][]{chap7_sch11} or never formed.  
The surficial Rhean Blue Pearls are easily erased over time, and are therefore geologically recent, but the apparent great age of the Iapetus ridge 
is such that large massive ring deposits of this kind would have been preserved through much of Saturn System history.  
The lack of prominent ridges on other Saturnian satellites is thus real and indicates that if any formed at those bodies, 
they formed very early and were erased in the accretional storm of projectile bombardment.

\subsection{Rings around Uranian moons}

Finally, the case for 
rings or ridges associated with
the Uranian moons is not clear, 
namely because Voyager lacked the color filters that Cassini observed the Rhea ring deposit with, 
but also because the 1986 Uranus encounter provided our only mapping to date.  
Like Voyager at Saturn, these maps are incomplete, and provided imaging on the outer Uranian satellites 
Titania and Oberon (where rings are likely to be more stable) at no better than 2 to 5 km/pixel, respectively, 
and only of the southern hemispheres.  
Thus it is likely that both Blue Pearls and an equatorial ridge on one or more of these moons were missed by 
Voyager, if they exist.
A return to the Uranian system will be required to determine if moon-rings ever formed there.

\section{Rings around Mars}

Studies of the giant planets have revealed a distinct connection between
small moons and dusty rings (see Chapter~13). The concept is simple---meteoroids impact
the surface of a moon and raise a cloud of dust. That dust escapes from
the moon's weak gravity but remains in orbit around the central planet.
For example, Jupiter's gossamer rings are associated with Amalthea and
Thebe 
\citep{chap7_bur99}. Similar rings emerge from several small
satellites of Saturn (e.g., Pan, Anthe, Aegaeon and Phoebe), Uranus
(Mab) and Neptune (Galatea). In an analogous way, we would expect
impacts into Phobos and Deimos to populate faint Martian dust rings.
\citet{chap7_sot71} was the first to predict the existence of rings of Mars,
using a similar argument. Since then, the hypothetical Martian rings
have been the topic of over thirty theoretical publications; see
\citet{KH97} for a historical summary.

Dynamical simulations by \citet{KH97}, building upon previous work \citep{JH95,KT95,Ishimoto96}, indicate that the Martian rings will have
some peculiar properties. Because of the strong influence of solar
radiation pressure, both rings are offset from the center of the planet,
with the Phobos ring displaced toward the Sun by $\sim$ one Martian radius,
whereas the Deimos ring should be displaced away from the Sun by several
radii. This peculiar result follows from the dynamics of individual dust
grains. Briefly, Mars' oblateness causes elliptical orbits of Phobos'
ring particles to precess faster than Mars' mean motion around the Sun,
whereas for Deimos, this precession is slower than Mars' mean motion.
Solar radiation pressure, which drives orbital eccentricities, is very
sensitive to the difference between the two motions; the results are
rings offset in opposite directions. The Phobos ring should be
equatorial and $\sim400$~km thick. The Deimos ring is predicted
to be much thicker, 10,000--15,000~km, and tilted out of the equatorial
plane toward the ecliptic by $\sim 15^{^\circ}$. The larger thickness and
tilt are both due to solar radiation pressure, which is a more effective
perturbation for this more distant ring. Similar solar perturbations are
responsible for the warps in Saturn's E ring, and for the tilt and thickness of the Phoebe ring (see Chapter~13). Also, the strong solar
radiation pressure quickly drives micron-sized grains out of orbit, by
inducing eccentricities large enough that they strike the planet; this 
leaves behind a ring that should be composed primarily of particles tens
of microns or larger \citep{Ham96,KH97}.

So far, attempts to observe rings around Mars have been unsuccessful.
\citet{chap7_dux88} used Viking images to put an upper limit on
the ring's normal optical depth $\tau < 3\times 10^{-5}$.
\textit{In~situ} observations by Mars-orbiting spacecraft of anomalies
in the solar wind magnetic field were interpreted in the 1980s as being
due to Martian rings, but more extensive measurements by the
magnetometer aboard \textit{Mars Global Surveyor} showed that observable
fluctuations are likely due to well-known solar wind or bowshock
phenomena \citep{Oieroset10}.

Earth-based, telescopic detection of rings around Mars is more difficult
than for analogous rings of the gas giants, because of the former's lack
of atmospheric methane. Methane has very strong absorption bands at,
e.g., 2.2~$\mu$m, so the the brightness of the giant planets drops
substantially at selected wavelengths, facilitating the detection of
faint rings. Mars has no analogous absorption bands. Nevertheless, the
Hubble Space Telescope (HST) has been used four times to search for
Martian rings. Two early attempts (HST programs GO-5493 and GTO-7176)
used inappropriate viewing geometries and did not succeed; these results
are unpublished. On May 28, 2001, Mars' hypothetical ring plane appeared
edge-on to Earth within weeks of its opposition, providing the best
Earth-based opportunity to detect these rings for several decades. Using
the Wide Field/Planetary Camera 2 (WFPC2), \citet{SHN06} obtained upper
limits of $\tau < 3\times 10^{-8}$ for the Phobos ring and $\tau <
10^{-7}$ for the Deimos ring. This limit was sufficient to rule out
rings at the upper end of the dust density predictions by \citet{KH97}.
A final attempt with HST, by the same team, employed a slightly
inferior viewing opportunity in December 2007. Using the finer
sensitivity of the Advanced Camera for Surveys (ACS), the observing plan
had the potential to detect rings 30--100 times fainter than the
previous limit. However, due to the failure of the ACS prior to the
observations, the system was again imaged using WFPC2 and the ring
detection threshold could not be improved.

Despite the lack of detections, the dynamics of dust in the Martian
environment is well understood, and there remains little doubt that dust
rings, at some very low level, must be present. The most plausible
remaining method for detecting them would entail placing a sensitive
dust detector into orbit around Mars. Japan's Nozomi spacecraft did
carry such an instrument, but it failed to enter orbit around Mars in
1999 as planned. Perhaps some future Mars mission will finally reveal
these long-sought rings.

\section{Rings around Pluto}

The discoveries of Pluto's small moons Nix and Hydra in 2005 
\citep{chap7_wea06} raised the possibility that it, like Mars, could harbor a
tenuous ring system. Charon is
less likely to be a major source of dusty rings because its
gravitational field will more efficiently retain any dust ejected from
its surface. Using the discovery images from HST, \citet{StefflStern07}
searched for rings in the orbits of the two small moons. They obtained
upper limits of a few $\times 10^{-7}$ in reflectivity which, depending
on the rings' albedos, corresponds to $\tau \lesssim 10^{-6}$. This is
comparable to the conservatively estimated $\tau < 10^{-6}$ suggested by
\citet{SternPluto06}.

The limiting factor in the search by \citet{StefflStern07} was the
extensive glare from Pluto and Charon. \citet{ShowCBET11} used an
alternative technique with HST to control for and subtract the glare
pattern, making a more sensitive ring search possible. That program did
not detect any rings, setting a new upper limit of $\tau <$ a few
$\times 10^{-7}$. However, it did reveal a fourth moon, Kerberos. The
following year, a more extensive HST observing program revealed a fifth
moon, Styx 
\citep{chap7_sho9253}.

With New Horizons en route to its July 2015 flyby of Pluto, the
revelation of such an extensive satellite system raised concerns about a
possible dust hazard to the spacecraft. However, dynamical studies in
advance of the flyby tended to minimize that risk. 
\citet{chap7_sho15} showed that the satellite system is on the edge of
chaos, which reduces the likelihood of a stable ring system. 
\citet{chap7_por15} showed that relatively few stable orbits exist between the
four small outer moons, and those that do exist require a nonzero
inclination. Furthermore, it had been noted that, somewhat
counterintuitively, solar radiation pressure is an important
consideration in the Pluto system because, although the Sun is very far away,
Pluto's gravity is also quite weak. 

\citet{chap7_sho10}
showed that micron-size grains are quickly driven into orbits that collide with
Charon, leaving behind particles primarily larger than $\sim 25$~$\mu$m.
%
More detailed models showed that 
Nix and Hydra should be the main dust provider for broad and tenuous rings 
through micrometeoroid impacts. 
Combined effects of radiation pressure, collisions of the ejecta with the larger bodies 
Pluto and Charon and escape, however, seriously limit the optical depth of such rings.
\cite{pop11} estimated optical depths on the order of $10^{-7}$ for grains between
0.1 and 100 $\mu$m in size. 
On the other hand, \cite{pir13} showed that radiation pressure remove very rapidly
(on year-scales) particles with sizes around 1~$\mu$m, leaving rings with normal
optical depth on the order of $10^{-11}$.

An extensive survey of the Pluto system was conducted
throughout the New Horizons mission approach phase. No new moons and no rings were
detected, and the spacecraft passed through the system safely. The final
upper limit on the dust optical depth was $10^{-7}$ 
\citep{chap7_spe15}. After the flyby, the spacecraft conducted an outbound search
for faint rings at high phase angles. However, the analysis of that data
set is still underway and no results have been reported.

Stellar occultations also provide an opportunity to search for rings. Optically
thin rings would be generally undetectable by this method, but rings that are
narrow and/or dark could potentially show up in occultations before they are
imaged directly. Historically, this is how the rings of Uranus and the arcs of
Neptune were discovered. 
\citet{chap7_boi14} conducted one such search, setting
an upper limit of 30--100~m for the equivalent depth of a narrow ring (see Chapter~4 for the formal definition of ``equivalent depth,'' a form of radially-integrated
optical depth). Similarly, 
\citet{chap7_thr15} set a limit of $\sim 170$~m assuming a nominal
ring of width 2.4 km.

\section{Rings around exoplanets}
\index{Exoplanets, rings around}

Studies of transient photometric changes have predominantly focused on objects that brighten, 
such as novae or gravitationally micro-lensed objects.  
The exceptions to this rule are the recent searches for exoplanet transits that  find periodic but shallow and short dimming events.
Thousands of exoplanets have been discovered using transit searches, however, with one exception, J1407
\citep{chap7_kenmam15}, 
these have not been interpreted in terms of exoplanetary ring systems.  
Detection algorithms that rely on periodicity,  fitting a known transit shape to the dimming event or that ignore deep 
events from eclipsing binaries would likely discard an eclipse from a  circumsecondary and circumplanetary disk 
\citep{chap7_mam12,chap7_men14,chap7_zul15}. 
Such eclipses could be rare  or have long periods  and if periodicity is required for a search
they would not be found
\citep{chap7_qui14,chap7_pet15}.
They could have an unusual light curve transit shape, and so would not be well fit by a planet transit or eclipsing 
binary light curve model
\citep{chap7_men14,chap7_don14,chap7_rat15}. 
The transits  could be deeper than expected for a planet transit and so would be classified as a possible 
eclipsing binary and so ignored in planet transit searches. 

\begin{figure}
\figurebox{80mm}{}{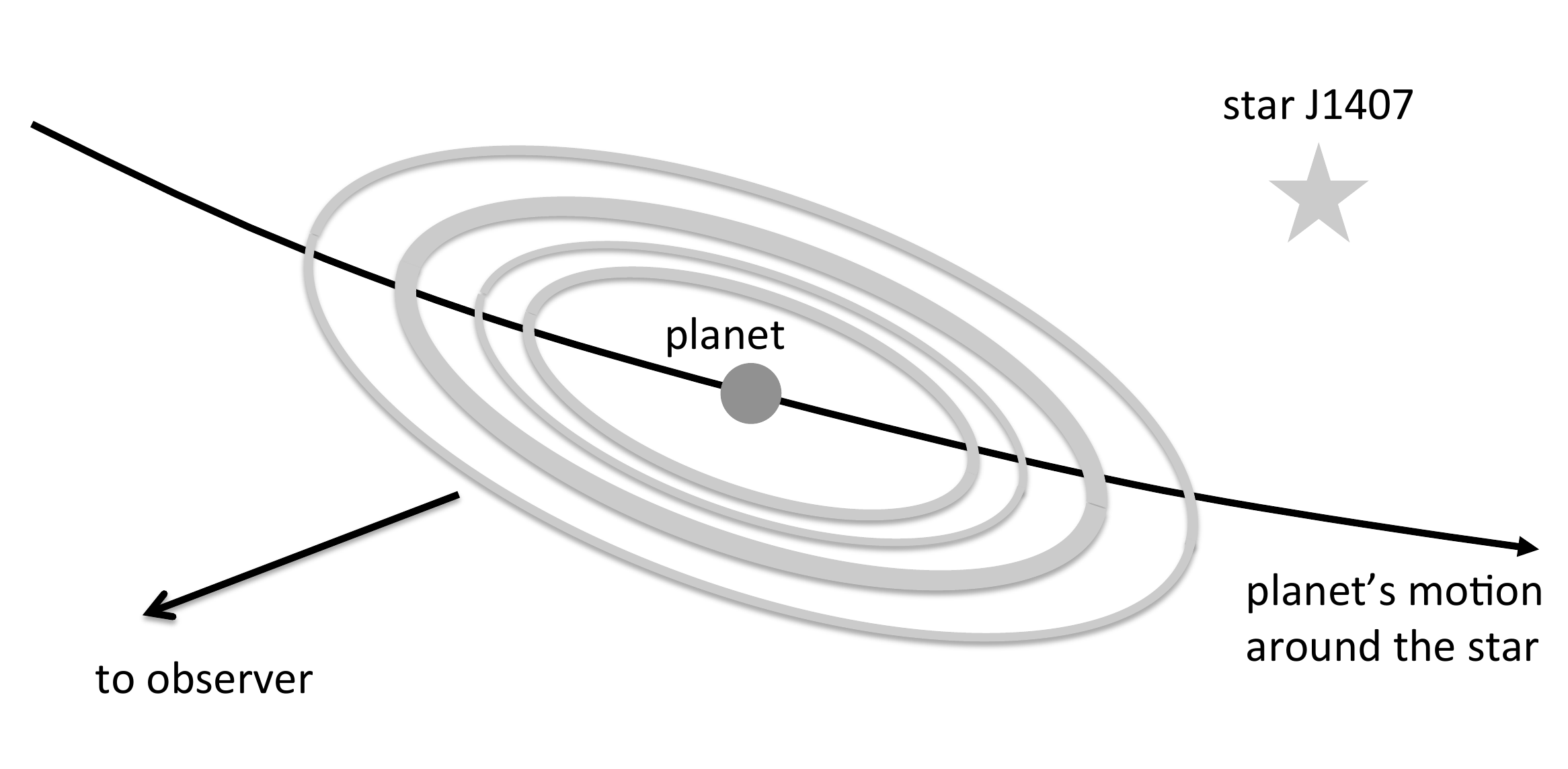}
\caption[Detecting an exoplanetary ring system]{%
Sketch illustrating the modeling of possible rings 
around J1407b,  the putative companion of J1407 (see text).
During its motion around the star, an exoplanet surrounded by rings 
may cause the dimming of the star as seen by an observer 
(here far away on the lower left side of the figure).
The sizes of the planet and its ring system are not on scale,
they have been enlarged for better viewing. 
}%
\label{chap7_fig_exo_ring}
\end{figure}

During an occultation (Fig.~\ref{chap7_fig_exo_ring}), 
structure in an occulting disk can be measured on the scale of a stellar radius or 
$\sim$~0.01 AU 
(about one million km),  a scale that is difficult or impossible to resolve directly.  
So if eclipsing exoplanetary ring systems are evident in light curves, they would allow detailed study of ring structure, 
potentially even giving constraints on their composition through time resolved spectroscopy or  from their wavelength 
dependent reflection and transmission properties.
%
%

Ring system outer radii are larger than planetary radii. 
The probability of detecting an extended object in transit depends on 
the fractional 
solid angle of the object covered in its orbit about the host star.

A spherical object of diameter $d$ object in orbit with semi-major 
axis a covers a band of area $2\pi d a$ during one revolution, 
i.e. a solid angle of  $2\pi d/a$ as seen from the star.
An observer viewing the star with line of sight
intersecting this band would see an eclipse of the star.
The todal solid angle covered by a sphere  being $4 \pi$,  
the ratio $p = (2\pi d/a)/(4\pi)= d/(2a)$ gives the probability that an eclipse is seen from 
a distribution of systems during a survey that goes on long enough to include the period of the orbit,
taking into account all possible orientations and assuming a uniform distribution of orientations.
%
%
If disks fill a large fraction of a secondary's Hill radius, 
then the probability that a sample of young stars star exhibits eclipses 
could reach as high as $10^{-4}$ (see equation 14  of \citealt{chap7_mam12}).

With the discovery of the J1407 dimming event, it was argued 
from the number above
that the probability of finding a dimming event  by an eclipse  
of a circumsecondary or exoplanetary ring system in a large photometric survey is not  small  \citep{chap7_mam12}. 
A large uncertainty affecting the detection probability for eclipsing disks is the lifetime of circumplanetary and circumsecondary 
disks, with younger stars being most likely to host extended and dense  circumplanetary or circumsecondary disks, 
and suggesting that young stellar populations would be most likely to exhibit disk or ring system eclipses. 

 The low fraction of field eclipsing binaries that exhibit disk like features suggests that the size and optical depth 
of ring systems and circumbinary disks decreases with age  \citep{chap7_men14}.
A survey of 40,000 stars in the 2MASS calibration database that  searched for dimming events and did not restrict 
the search to periodic light curves, primarily found new eclipsing binaries but also found dimming events 
from variable young stars \citep{chap7_qui14}. 
The KELT photometric survey recently discovered long and deep
dimming events from young stars.
The 2014 dimming of RW Aurigae A was greater
than 2 magnitudes and lasted 6 months \citep{chap7_pet15}.
The star RW Aur also exhibited a 6 month long 2 mag dip in 2010 \citep{chap7_rod13}, while
DM Ori exhibited, twice, dips of 1.5 mag lasting 6 months \citep{chap7_rod16}.
These types of events are sometimes called ``dippers".
Even though the Kepler mission discovered thousands of short period planets in transit, 
none so far exhibit ring systems -- perhaps short period planets cannot support extended ring systems \citep{chap7_hed15}. 
Because they did not restrict their search to periodic systems or light curves that brighten, 
the citizen scientists in the Planet Hunters consortium recently discovered an F star, KIC 8462852, 
exhibiting irregularly shaped, aperiodic dips in flux down to below the 20\% level \citep{chap7_boy16}.
Surprisingly this star does not exhibit the behavior of  a young star.  The dips in the light curve might be explained 
with a dust cloud created by a destructive impact between two large planetesimals or families of evaporating comets
\citep{chap7_bod16a}.
Another star, the white dwarf WD1145+017 exhibits 3-12 minute deep (0.5 mag) transit
events (or dips) that could be due to disintegrating comets \citep{chap7_gan16}.
Recently more than twenty young ($\sim$~10 Myr old) late-K and M dwarf stars were observed in the 
Kepler Mission K2 Campaign Field 2 that host protoplanetary disks and exhibit 
quasi-periodic or aperiodic dippers \citep{chap7_ans16}.
Magnetospheric truncation and accretion models can explain why dusty material is lifted out of the 
midplane to obscure the star causing the light curve dips and why so many young low mass stars are dippers 
\citep{chap7_bod16b}.  

\begin{table*}
\caption{Eclipsing Circumsecondary Disks}
\begin{tabular}{lllllll}
\toprule%
Object                                              & Eclipse length & Period & Depth$^{a}$  & Primary & Disk Temp. \\ 
\hline
EE Cep$^{b}$                                & 30-60 days    & 5.6 yr           & 0.6-2.1     & B5eIII  & 900K    \\ 
Epsilon Aurigae$^{c}$                  & 2 years           & 27.1 yr        & 0.8           & F0 I      & n.a.   \\ 
OGLE-LMC-ECL-17782$^{d}$    &  2 days           & 13.35 days & 0.5           & B2       & 1200K \\Ê
OGLE-LMC-ECL-11893$^{e}$    &  15 days         & 468 days    & 1.5           & B9       & 6000K  \\ 
OGLE-BLG 182.1.162852$^{f}$ & 100 days    & 3.5 years        & 1--2          & n.a.     & 300 K  \\  
J1407$^{g}$                                    &  60 days      & $>$4 years   & 4              & K5V    & 200K   \\ 
\botrule
\end{tabular}
\begin{tabnote}
$^{a}$Measured in magnitude drop.
$^{b}$\citet{chap7_gal12}. 
$^{c}$\citet{chap7_cha11}. 
$^{d}$\citet{chap7_men14}.  
$^{e}$\citet{chap7_don14}. 
$^{f}$\citet{chap7_rat15}. 
$^{g}$\citet{chap7_ken15}. 
\end{tabnote}
\label{chap7_tab_circumdisk}
\end{table*}

\subsection{Notes on individual objects}

Two bright stars, 
EE-Cep and Epsilon Aurigae,  have long been known to exhibit deep eclipses.
These are  both long period systems hosting circumsecondary eclipsing disks with early type primary stars.
Both disks have radial structure such as a central clearing and in EE-Cep this  causes asymmetry in the light curve 
(e.g., \citealt{chap7_gal12}).
Re-examination of eclipsing binary light curves in archival data have revealed three more eclipsing disk systems, 
OGLE-LMC-ECL-11893 with a 468 day period \citep{chap7_don14} and OGLE-BLG 182.1.162852, 
a bulge object with a 3.5 year period \citep{chap7_rat15} and OGLE-LMC-ECL-17782, 
exhibiting 2 day eclipses in a 13 day period; this likely host a transient B-star blow-out disk 
\citep{chap7_men14}. 
Each eclipse of OGLE-LMC-ECL-11893 is remarkably similar and multi-color photometry shows that
dust in the disk causes reddening \citep{chap7_don14}. 
The eclipse shape can be fit with either  a thin dusty disk or a thick gas and dust disk \citep{chap7_sco14}. 
Existing multicolor photometric observations could in the future be used to study the dust properties.
These known eclipsing disk systems are listed in Table~\ref{chap7_tab_circumdisk}.  Disk temperatures are estimated
from the orbital period and the luminosity of the primary and span a wide range suggesting that
these systems may in future provide interesting settings to study disk composition through spectroscopy. 

An interesting case is
J1407 (1SWASP J140747.93-394542.6), 
a 16 million year old,
pre-main sequence K5-type star of some  0.9 solar mass
in the Sco-Cen OB association. 
It exhibited a complex 54 day deep eclipse in April 2007,
with a maximum depth greater than 3 magnitudes 
\citep{chap7_mam12,chap7_van14}. 
The long eclipse was discovered in a Super Wide Angle Search for Planets (SuperWASP)  light curve
but a few data points from the All Sky Automated Survey  (ASAS) confirmed that the star dropped in brightness in 2007.
Continued monitoring and high contrast imaging rule out a bright or stellar secondary object \citep{chap7_ken15}.
An optically thick ring passing in front of the star, causes a change in slope (flux variation per unit time) 
dependent on the  angular rotation rate of the ring (also see limits on radius of occulting objects 
in the KIC 8462852 system by \citealt{chap7_boy16}).

More detailed modelings of the slopes have been conducted by 
\cite{chap7_kenmam15} and \cite{chap7_ken15}, see Fig.~\ref{chap7_fig_exo_ring}.
Using slope changes in the light curve, each corresponding to a ring edge, the J1407 eclipsing system
has been modeled as a complex set of more that 35 thin rings lying in a oblique plane.
The large slopes at some epochs suggest that the secondary object hosting the ring system
orbits at no more than a few AU from the primary star on an eccentric orbit,
with a period estimated from a few to some 30 years.
The ringed object would be a giant planet of some 15-25 Jovian masses, 
while the ring system would contain about one Earth mass and
span a diameter of about 180 millions km (1.2~AU).
This can be compared to Saturn's rings, which contain about $10^{-5}$ Earth mass
and is more than 600 times smaller than the system considered here.

The complex substructure suggests that the ring system is very thin and hosts moons
that maintain sharp edges at Lindblad resonances, or open gaps in the disk. 
Crude estimations based on the width of one of these gaps suggest that it could stem 
from a Mars-size or small-Earth type object \citep{chap7_kenmam15}.
As more constraints of the disk thickness and planet mass are gathered,
Eqs.~\ref{chap7_eq_shepherd_size} and \ref{chap7_eq_gap_opener} may be used
to better assess the masses of those putative moons.

As a word of caution, 
we note that continued photometric monitoring of J1407 (Erin Scott, private communication) 
has not revealed new eclipse episodes, as is expected if the ringed planet has an
orbital period of a few years.
If  ongoing monitoring fails to find new eclipses over the next decades, 
then it may become impossible to account for the eclipse 
with an extended ring system orbiting a secondary object.

This said, an intringuing issue is the fact that the J1407 
putative ring system would extend much beyond the planet Roche limit 
(usually some 2-3 planetary radii).
These rings would then represent transient features en route towards an
accretion process that will form a retinue of moons around the planet.
Scaling from models of the proto-Jovian nebula, \cite{chap7_mam12}
estimate that the lifetime of a circum-Jovian disk could be as long as 
several millions years, thus comparable to the age of J1407.
However, this challenges the mainstream idea that rings
exist only inside the Roche limit of their central planets, as accretion should
proceed very rapidly (over a few orbital revolutions) to form moons.
To prevent such outcome, Toomre's parameter $Q$
should be maintained just above unity. 
In that context, Eq.~\ref{chap7_eq_toomre} can be re-written \citep{chap7_sic06}:
\begin{equation}
\frac{h}{a} \sim \frac{M_r}{M_p},
\label{chap7_eq_ring_mass}
\end{equation}
assuming a uniform ring of radius $a$, mass $M_r$ and thickness $h$
surrounding a planet of mass $M_p$.
For $M_r$ comparable to Earth's mass, $M_p$ of some 20~Jovian masses
and $a$ a little bit above 1~AU (see above), this implies $h \sim 30,000$~km.
At this point, the mechanism causing the stirring of the disk and
maintaining this thickness remains to be explained.

\section{Concluding remarks}

As shown in this chapter, 
rings beyond the giant planets appear to be more common features than previously thought.
This has important implications at different levels.

First, this raises the question of whether  
rings share some basic, universal physics, or, on the contrary,
if they follow a wide variety of disconnected behaviors depending on the context.
For instance, Chariklo's rings and the material surrounding J1407b
exhibit sharp edges or gaps, that are also encountered in Saturn's and Uranus' rings.
Are those features all caused by shepherding nearby bodies (satellites or planets), 
or do they stem from other, yet to be described physical processes?
In fact, none of the ``moonlets" that are thought be responsible for the narrow rings
or gaps in the Saturnian C ring and Cassini Division have been discovered so far
\citep{chap7_col09}.
As high resolution imaging is steadily improving thanks to larger telescopes,
adaptive optics or space instruments, it is now of paramount importance to discover
(or rule out) the presence of confining bodies associated with sharp edges and gaps
in the newly discovered rings.

At another level, 
rings can tell us a lot about the body they encircle.
With the advent of the European Space Agency GAIA mission,  
star catalogs with absolute accuracy of a fraction of milliarcsec will be released soon.
In that context, stellar occultations by Chariklo's rings will be routinely observed by many teams. 
Those campaigns will then provide accuracies of better than one kilometer on the rings' orbital elements. 
This might lead to the discovery of ring proper modes, as observed in Uranus' rings 
(see the Chapter 4 by Nicholson \textit{et al.}) and then provide the 
ring particles' mean motion, a direct way to determine Chariklo's mass, and thus
its density through its dimensions.
In the same vein, the rings' precession rates could yield
Chariklo's dynamical oblateness $J_2$, an important parameter to understand its 
internal structure.
In short, rings may be precious probes of the gravity field of their host body.
 
Those programs are not restricted to Chariklo, but also aimed at searching for material
around other Centaurs, TNOs and asteroids. 
This may lead to the discovery of new ring systems, or rule them out with a safe margin.
It will then be possible to address on firmer ground the question of whether ``small body" rings 
exist only around Centaurs, and why it is so, 
or if on the contrary, they are also present around very remote TNOs or even nearby asteroids.
If exclusive to Centaurs, rings could be 
the witnesses of  the troubled history of those objects (e.g. stemming from close encounters with giant planets), or 
a mere endogenous product associated with cometary activities of  those bodies, or
the outcome of a fine tuning of icy composition, size and temperature conditions, or
the result of some other unkown processes.

As discussed in this chapter, 
rings might also have existed around the Saturnian satellite Iapetus and Rhea.
However, those rings should have been quite different from each other, with a massive 
(relative to the satellite) disk that fell along Iapetus' equator early in the history of the Solar System, 
and a relatively recent low-mass ring that sprinkled Rhea's equator.
In any case, they would be remnants of interesting processes, and would tell
us how a collisional disk or an object can be driven toward a body through tidal interactions
and fall onto its equator.
In that vein, it is important to check -- using well-sampled stellar occultations --
if  a ridge exists along Chariklo's equator (or around other Centaurs). 
This would be a nice confirmation that rings may indeed explain Iapetus' equatorial feature.

Turning to exoplanetary rings, we note that
in 2012 transient events were considered uninteresting and completely ignored. 
The discovery of new eclipsing circumsecondary disks \citep{chap7_don14,chap7_rat15},  
a candidate exoplanetary ring system \citep{chap7_mam12,chap7_kenmam15} 
and  deep transient dimming events in both young and old stars from Kepler Mission data 
\citep{chap7_boy16,chap7_ans16} 
imply that photometric observations can uncover new eclipsing disk systems.

Up to now all eclipsing and transient dimming events have been found in archival data, 
making it difficult to follow up non-periodic or long period eclipsing systems.  
In two cases, the same dimming events were found in more than one photometric archive, 
giving confirmation (J1407, \citealt{chap7_mam12} and the dimming of V409 Tau, \citealt{chap7_rod13}). 
Some of the dimming events seen in  KIC 8462852 could have been detected from the ground.
This system is now being monitored for new dimming events which may allow a multicolor photometric study. 
It is possible to mount a transient detection program that triggers on dimming events allowing
multicolor or high cadence observations (and possibly spectroscopic observations) of rare, 
and long period, dimming events.

Future and more accurate photometric studies of larger populations of stars could detect dimmings 
caused by an exoplanetary ring system as rich and old as Saturn's as well as counterparts at earlier epochs.

In that context, an interesting issue is the location of those rings relative to the exoplanet's Roche limit.
While Chariklo's rings seem  to lie a little bit outside of, but still near 
Chariklo's Roche limit, the putative exoplanetary rings associated with J1407b are 
well outside that range. This undermines the paradigm of rings as collisional
disks residing inside the Roche limit of the central body in order to
prevent rapid accretion into individual objects.
So, 
we are either very lucky to observe today the J1407b ring system before it coalesces into satellites, 
or our understanding of accretion time scales needs revisions. 
In any case, an estimation of the probability to detect by mere chance a ring system among all the
transit events now observed is very much wanted.
This might put constraints on the efficiency of confining mechanisms and 
on accretion time scales, thus allowing us to better understand the various steps
that led to the formation of planets and satellites, including in our own Solar System.

\vspace{10pt}
B.S. acknowledges funding 
from the French grant ``Beyond Neptune II" (ANR-11-IS56-0002) and
from the European Research Council under the 
European Community's H2020 (2014-2020/ ERC Grant Agreement no. 669416 ``LUCKY STAR").
K.J.W. acknowledges funding from the NASA Origins program, and NASA SSERVI program 
(Institute of the Science of Exploration Targets) through institute grant number NNA14AB03A.


\printindex

\end{document}